\documentclass[onecolumn,showpacs,preprintnumbers,amsmath,amssymb,floatfix,nofootinbib,showkeys]{revtex4}

\usepackage{graphicx}

\usepackage{bm}
\usepackage{amsfonts}
\usepackage{lineno,hyperref}
\usepackage{array}
\usepackage{microtype}
\usepackage{float}
\usepackage{epstopdf}
\usepackage{mathrsfs}

\usepackage{color}
\usepackage[utf8x]{inputenc}
\SetUnicodeOption{combine}
\usepackage[english]{babel} % 'french', 'german', 'spanish', 'danish', etc.
\usepackage{amssymb}
\usepackage{txfonts}
\usepackage{makecell}
\usepackage{subfigure} % in preamble
\usepackage{esint}
\usepackage[figurename=FIG.]{caption}
\usepackage[labelsep=period]{caption}
\begin{document}
{\centering
	\makebox[\linewidth]{
		\includegraphics[width=1\linewidth]{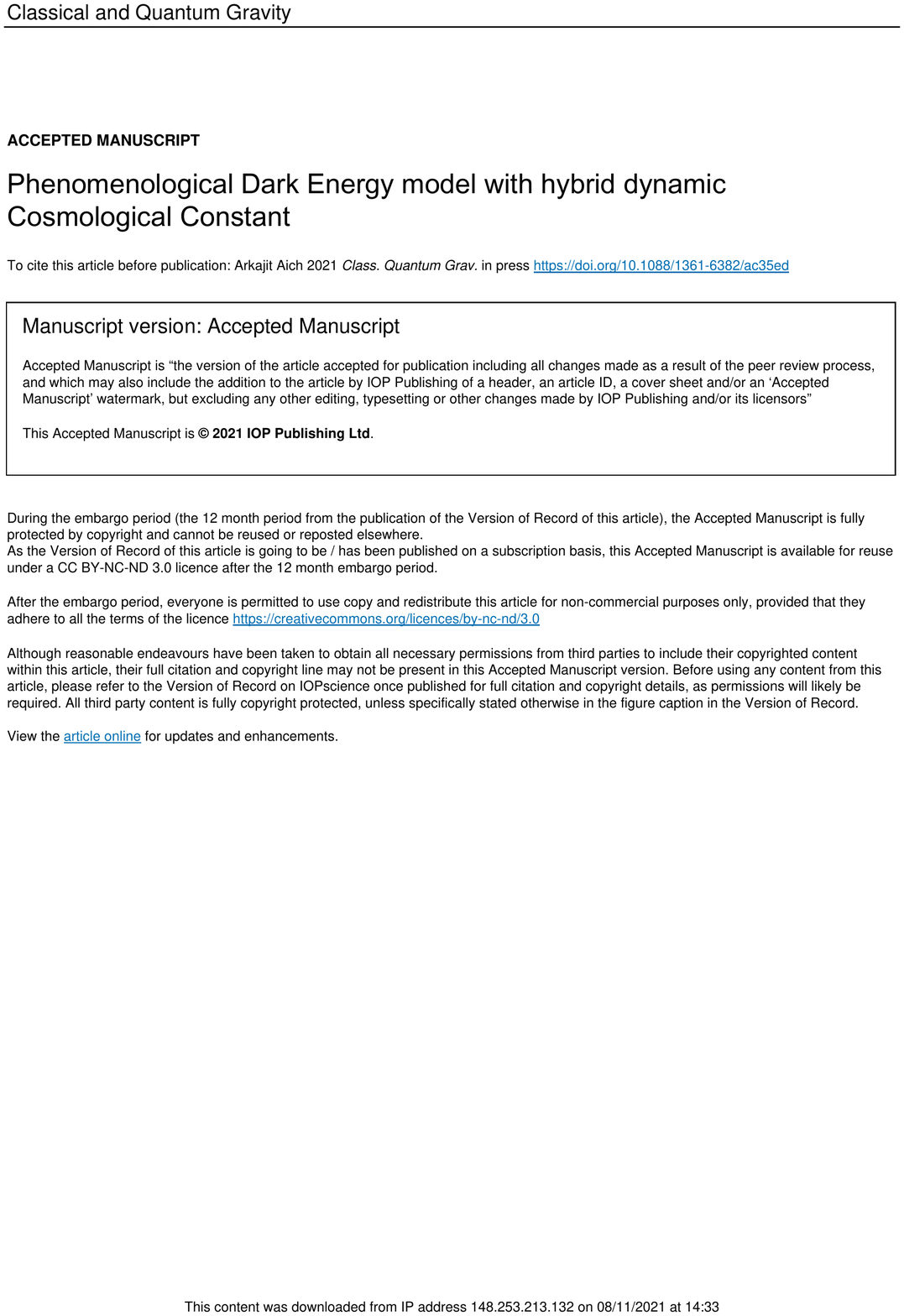}}}
\title{Phenomenological Dark Energy model with hybrid dynamic Cosmological Constant}

\author{Arkajit Aich}
\email{arkajit.aich.kolkata@gmail.com}
\address{School of Computing, University of Portsmouth, University House, Winston Churchill Ave, Portsmouth PO1 2UP, United Kingdom}

\begin{abstract}
We investigate Dark Energy by associating it with vacuum energy or Cosmological constant \textit{$\mathit{\Lambda}$} which is taken to be dynamic in nature. Our approach is phenomenological and falls within the domain of variable-\textit{$\mathit{\Lambda}$} Cosmology. However, motivated by quantum theory of metastable vacuum decay, we proposed a new phenomenological decay law of \textit{$\mathit{\Lambda}$}(\textit{t}) where \textit{$\mathit{\Lambda}$}\textit{(t)} is a superposition of constant and variable components viz. \textit{$\mathit{\Lambda}$}(\textit{t}) =  \textit{$\mathit{\Lambda_{C}}$} + \textit{$\mathit{\Lambda_{v}}$} which is indicated by the word $``$hybrid dynamic$"$  in the title. By taking a simplified two-fluid scenario with the Universe consisting of Dark Energy and another major component, we found the solutions for three particular phenomenological expressions and made a parametrization of the model in terms of dilution parameter (the dilution parameter has been defined in the text as the exponent of scale factor in the expression of density of the other major component, representing the dilution of the component with the expansion of Universe in the presence of dynamic Dark Energy). For pressureless Dust and dynamic Dark Energy Universe, we found the present-day matter density (\textit{$\mathit{\Omega_{m0}}$}) and dilution parameter (\textit{u}) to be \textit{$\mathit{\Omega_{m0}}$} = \(0.29 \pm 0.03\),  \( u~= 2.90   \pm  0.54 \) at 1\textit{$\mathit{\sigma}$} by analysing 580 supernova from Union 2.1 catalogue. The physical features of the model in regard to scale factor evolution, deceleration parameter, cosmic age has also been studied and parallels have been drawn with \textit{$\mathit{\Lambda}$}CDM model. The status of Cosmological problems in the model has also been checked which showed that the model solves the Cosmological Constant Problem but the Coincidence problem still exists in the model. 


\end{abstract}

\maketitle
	
\section{Introduction} \label{sec1}

After developing the General Theory of Relativity (GTR)~\cite{Einstein1}, which is a geometric theory of Gravitation, Einstein attempted to build a Cosmological model based on GTR. In order to keep the model Universe static, Einstein added an ad-hoc constant ~\cite{Einstein2} namely, the Cosmological Constant \textit{$\mathit{\Lambda}$} representing Universal repulsion in GTR. However, Edwin Hubble’s milestone discovery in 1929 that the Universe is expanding~\cite{Hubble} lead to discarding of a static Universe concept. Consequently, Friedmann-Lemaitre-Robertson-Walker (FLRW) family of models, which was built upon a Cosmological principle-based solution of GTR (FLRW metric)~\cite{Lemaitre} and represented an expanding Universe, became established as the standard Cosmological models of the era. The status quo was disrupted towards the end of 20th century when Supernova Cosmology project team and high-Z supernova search team jointly inferred from their observational studies ~\cite{Perlmutter,Riess} that the Universe is expanding at an accelerating rate and the accelerated expansion could not be explained within the framework of standard FLRW Cosmology. In order to have a physical basis for the observed acceleration, a hypothetical unknown entity, namely, Dark Energy (DE) has been assumed to exist in the Universe which is responsible for the observed acceleration. Since then, one of the fundamental quests of modern Cosmology has been to solve the mystery of DE. 

The exact nature of Dark Energy is still unknown. Present investigation into the Dark Energy problem revolves around building and studying DE-based Cosmological models by choosing various probable candidates to represent it. Among them, the historical \textit{$\mathit{ \Lambda}$} term of Einstein has  been resurrected and is also studied as a possible candidate for Dark Energy ~\cite{Demianski}. The current standard model of Cosmology viz. base-Lambda Cold Dark Matter (\textit{$\mathit{\Lambda}$}CDM) model takes up this approach where DE is represented by \textit{$\mathit{\Lambda}$}. In this line of investigation, \textit{$\mathit{\Lambda}$} is usually interpreted as vacuum energy arising from Quantum fluctuations. However, there are two fine-tuning issues associated with it ~\cite{Cheng}: (i) the Cosmological Constant problem  - The estimated theoretical value of quantum vacuum energy (\textit{$\mathit{\Omega}$\textsubscript{vac}}) at Planck scale and present observed values of Cosmological Constant energy density (\textit{$\mathit{\Omega}$\textsubscript{$\Lambda$}}) has a discrepancy of about 120 orders of magnitude which indicates there must be some unknown mechanism that fine-tunes the value of \textit{$\mathit{\Lambda}$} to its present observed value., (ii) the Coincidence problem  - Observational results show that matter and vacuum density are nearly equal in present epoch despite scaling differently. Therefore, the initial conditions must be fine-tuned to achieve this. 

A solution to these issues is to assume that \textit{$\mathit{\Lambda}$} is a time-varying (decaying) parameter instead of a constant i.e. \textit{$\mathit{\Lambda}$} has decreased from its initial large value to its present small value which can address the issues. Following the argument of ~\cite{Overduin1}, variation of \textit{$\mathit{\Lambda}$} is possible within general relativity if we define an effective momentum tensor \textit{$T^{eff}$} = \textit{$T^{\mu \nu}$} - (\textit{$\mathit{\Lambda}$}/8\textit{$\mathit{\pi}$}\textit{G})\textit{$g^{\mu \nu}$} and assume that \textit{$T^{eff}$} satifies energy conservation. However, most analysis done in this area do not attempt to derive such a dynamic \textit{$\mathit{\Lambda}$} term from any fundamental theory and simply assumes an expression of \textit{$\mathit{\Lambda}$}(\textit{t}). They are classified as Phenomenological models of dynamic \textit{$\mathit{\Lambda}$} which are also important contenders for Dark Energy. According to Sahni and Starobinsky~\cite{Sahni} phenomenological dynamic \textit{$\mathit{\Lambda}$} models can be classified into three categories based on $``$fundamentality$"$ : (i) Kinematic, (ii) hydrodynamic, (iii) field theoretic. In Kinematic models, dynamic expression of \textit{$\mathit{\Lambda}$} is justified from dimensional arguments whereas in hydrodynamic models the dynamic \textit{$\mathit{\Lambda}$} term is estimated by associating it with a barotropic fluid. In field theoretic models, dynamic \textit{$\mathit{\Lambda}$} term is associated with a new physical classical field for which the authors coined the term $``$Lambda field$"$ .  A list of various such \textit{$\mathit{\Lambda}$} decay models which are used in literature, irrespective of categories described above has been listed in ~\cite{Overduin1}. In this regard, it should be mentioned here that some of these works dates back before the discovery of acceleration of the Universe when Dark Energy was not even in the picture. In fact, the first proposal of a time varying \textit{$\mathit{\Lambda}$} came as early as 1933 through the works of Bronstein ~\cite{Bronstein}. Several works with variable \textit{$\mathit{\Lambda}$} were carried out in the eighties ~\cite{Lima1}\footnote{check the review ~\cite{Lima1} and references within for an account of various works done with time varying Cosmological Constant} way before discovery of acceleration of Universe. In these older works therefore, time-varying \textit{$\mathit{\Lambda}$} does not represent Dark Energy, rather most of the works were motivated with the intentions of solving different issues of standard Cosmological models. We will however stick with the modern recipe of associating dynamic \textit{$\mathit{\Lambda}$} term with Dark Energy. 

On physical grounds, time-variation of \textit{$\mathit{\Lambda}$} indicates the quantum mechanical process of vacuum/dark energy decay which ideally must be justified from quantum mechanical principles. In fact, a quantum approach~\cite{Szyd} proposes that Dark or Vacuum Energy is in a metastable state (false vacuum) and is decaying towards a stable state (true vacuum) as cosmic time \( t \rightarrow \infty \). This provides a physical basis for time-variation of \textit{$\mathit{\Lambda}$} but in most phenomenological \textit{$\mathit{\Lambda}$} decay laws available in literature, vacuum Energy density approaches zero as  \( t \rightarrow  \infty \) and not to a stable value.  In an attempt to connect the quantum approach to phenomenological approach, we will introduce a new type of phenomenological decay law in this work. Instead of going in the traditional route, we will assume a hybrid\footnote{The terms $``$hybrid dynamic$"$  and $``$pure dynamic$"$  are used throughout this text to refer to the cases of variation of $ \Lambda $\ with the additive constant and without the additive constant, respectively.  These phrases have no definite mathematical meaning and are only used to conveniently differentiate between the two types of variation of $ \Lambda$.\ } dynamic nature of  \textit{$\mathit{\Lambda}$}(\textit{t})  and propose  \textit{$\mathit{\Lambda}$}(\textit{t}) is a superposition of a constant component and a time varying component. Mathematically,  \textit{$\mathit{\Lambda}$}(\textit{t}) =  \textit{$\mathit{\Lambda_{C}}$} + \textit{$\mathit{\Lambda_{v}}$}   (where  \textit{$\mathit{\Lambda_{C}}$}  and  \textit{$\mathit{\Lambda_{v}}$} refers to the constant and variable component of \textit{$\mathit{\Lambda}$} respectively) with \textit{$\mathit{\Lambda}$}(\textit{t}) \(\rightarrow\) \textit{$\mathit{\Lambda_{C}}$}  as  \( t \rightarrow  \infty \) replicating the metastable vacuum decay scenario from a phenomenological approach. This type of hybrid dynamic nature of \textit{$\mathit{\Lambda}$} keeps the main spirit of phenomenological dynamic \textit{$\mathit{\Lambda}$} models intact since \textit{$\mathit{\Lambda}$} is still a decaying parameter but it also adds a new dimension to phenomenological variable \textit{$\mathit{\Lambda}$}-Cosmology by linking it to quantum mechanical principles. An added advantage of this type of model is that standard \textit{$\mathit{\Lambda}$}CDM model and pure\footnote{see footnote 2} dynamic \textit{$\mathit{\Lambda}$} models can be readily recovered as special cases simply by setting  \textit{$\mathit{\Lambda_{C}}$}  or  \textit{$\mathit{\Lambda_{v}}$}  equal to zero with suitable choice of parameters. This flexibility provides ground for comparison with the standard model when confronted with observations. Another important feature of this type of model is that the presence of the additive constant in the expression of \textit{$\mathit{\Lambda}$} ensures the signature flip of deceleration parameter (\textit{q}) which is not readily obtained in pure dynamic \textit{$\mathit{\Lambda}$} models. So, the hybrid \textit{$\mathit{\Lambda}$} model has certain advantages, and, in this work, we will explore this model.

\section{MATHEMATICAL FORMULATION} \label{sec2}

A realistic Universe is made up of multiple components but in order to keep the calculations simple, we will assume the Universe to be made up of two fluids - exotic (decaying) Dark Energy component represented by phenomenological hybrid dynamic \textit{$\mathit{\Lambda}$} and another major component of the Universe, besides Dark Energy/vacuum. Furthermore, we will also assume a flat Universe in line with WMAP results ~\cite{Hinshaw}. 

The Einstein Field Equation (EFE) of GR including a time-varying \textit{$\mathit{\Lambda}$} term takes the form,

\begin{equation}\label{eq1}
G^{\mu \nu }=\ -8\pi G\left[T^{\mu \nu }-\ \frac{\mathit{\Lambda}\left(t\right)}{8\pi G}g^{\mu \nu }\right]\
\end{equation} 
(where we have used relativistic units  \( c = 1 \)  )

We will assume the Cosmological Principle to be valid even in the presence of a variable-\textit{$\mathit{\Lambda}$} term, therefore the background geometry of the Universe will follow the standard FLRW metric given by,

\begin{equation}\label{eq2}
{ds}^2=\ -{dt}^2+{a\left(t\right)}^2\left[\frac{{dr}^2}{1-kr^2}+\ r^2\left({d\theta }^2+{sin}^2\theta d{\varphi }^2\right)\right]\ 
\end{equation} 

The metric (2) and EFE (1) readily gives the Cosmological Field Equations for  \textit{$\mathit{\Lambda}$}(\textit{t}) 

\begin{equation}\label{eq3}
{\left(\frac{\dot{a}}{a}\right)}^2=\ \frac{8\pi G\rho }{3}+\ \frac{\mathit{\Lambda}\left(t\right)}{3}\ -\frac{k}{a^2}\
\end{equation} 

\begin{equation}\label{eq4}
\ddot{\frac{a}{a}}=\ -\frac{4\pi G}{3}\left(\rho +3P\right)+\ \frac{\mathit{\Lambda}\left(t\right)}{3}\
\end{equation}

Introducing the equation of state \textit{P = }\textit{$\mathit{\omega}$}\(\rho\) where \textit{$\mathit{\omega}$} denotes the Equation of state (EOS) parameter of the other dominating component, we can write (4) as,

\begin{equation}\label{eq5}
\ddot{\frac{a}{a}}=\ -\frac{4\pi G}{3}\left(1+3\omega \right)\rho +\ \frac{\mathit{\Lambda}\left(t\right)}{3}\
\end{equation}

The Energy conservation equation for Cosmology with variable \textit{c},\  \textit{$\mathit{\Lambda}$} and \textit{G} was derived by Vereshchagin and Yegorian ~\cite{Vereshchagin} which reduces to the following form when \textit{c} and \textit{G} is kept constant but \textit{$\mathit{\Lambda}$} is taken to be variable:

\begin{equation}\label{eq6}
\dot{\rho }\ +\ 3\frac{\dot{a}}{a}\left(\rho +P\right)=\ -\frac{\dot{\mathit{\Lambda}}}{8\pi G}\
\end{equation}

Since we are associating Cosmological Constant with vacuum energy fluid represented by \textit{P}\textsubscript{$\mathit{\Lambda}$}  = \textit{$\mathit{\omega}$}\textsubscript{$\mathit{\Lambda}$}\textit{$\mathit{\rho}$}\textsubscript{$\mathit{\Lambda}$}, therefore the usual relation between vacuum energy density and Cosmological Constant will be valid even when \textit{$\mathit{\Lambda}$} is a time-varying parameter. Henceforth, in general for \textit{$\mathit{\Lambda}$} we can write,

\begin{equation}\label{eq7}
{\rho }_{\mathit{\Lambda}}=\ \frac{\mathit{\Lambda}\left(t\right)}{8\pi G}
\end{equation}

In case of hybrid dynamic nature of \textit{$\mathit{\Lambda}$} undertaken in this work, equation (7) takes the specific form,

\begin{equation}\label{eq8}
{\rho }_{\mathit{\Lambda}} = \frac{{\mathit{\Lambda}}_C+\ {\mathit{\Lambda}}_v}{8\pi G}\
\end{equation}

An important aspect of Phenomenological dynamic \textit{$\mathit{\Lambda}$} models is that correspondence between models of different categories can often be established ~\cite{Sahni}. Therefore, it does not really matter which approach one takes, and, in this paper, we will take the kinematic approach where expressions of \textit{$\mathit{\Lambda}$} are justified from dimensional arguments. From the mathematical point of view, since our hybrid dynamic \textit{$\mathit{\Lambda}$} model just adds an additive constant to the variable \textit{$\mathit{\Lambda}$} term, therefore if we use the dimensionally valid expressions of \textit{$\mathit{\Lambda}$} available in literature as phenomenological expressions of \textit{$\mathit{\Lambda}$\textsubscript{v}}, it will still be valid on dimensional grounds. 

In particular, we will choose three such expressions of \textit{$\mathit{\Lambda}$} for the present work which are frequently used in literature ~\cite{Paul}\footnote{Check the references within ~\cite{Paul} for different works done with these three types of models.}:\textit{ }(i)  \(\mathit{\Lambda} ~\sim \left(\dot{a}/{a} \right) ^{2} \) ,\ \ \  (ii)  \(  \mathit{\Lambda} ~ \sim   \ddot{a}/a \)  ,\  (iii)  \(  \mathit{\Lambda}~\sim  8 \pi G \rho  \) 

We will use these expressions as expressions of  \textit{$\mathit{\Lambda_{v}}$}  with suitable proportionality constants and solve for each one of them separately to obtain the expression of density parameters.

\subsection{Solution for Phenomenological model \( ~ \mathit{\Lambda} _{v}=3 \alpha  \left( \frac{\dot{a}}{a} \right) ^{2} \)}
Using the ansatz  \(  \mathit{\Lambda} _{v}=3 \alpha  \left( \frac{\dot{a}}{a} \right) ^{2} \) , we have:

\begin{equation}\label{eq9}
\mathit{\Lambda}\left(t\right)={\mathit{\Lambda}}_c\ +3\alpha {\left(\frac{\dot{a}}{a}\right)}^2,\
\end{equation}

Substituting (9) in (3) and setting  \( k = 0 \)  for Flat Universe, we get:

\begin{equation}\label{eq10}
\frac{\mathit{\Lambda}\left(t\right)-{\mathit{\Lambda}}_C}{3\alpha }=\ \frac{8\pi G{\rho }}{3}+\ \frac{\mathit{\Lambda}\left(t\right)}{3}\
\end{equation}

Equation (10) reduces to 

\begin{equation}\label{eq11}
\mathit{\Lambda}  \left( t \right) =  \left( \frac{ \alpha }{1- \alpha } \right)  \left[ 8 \pi G \rho + \frac{ \mathit{\Lambda} _{C}}{ \alpha } \right] 
\end{equation}

Taking derivative of (11) w.r.t Cosmic time (\textit{t}), we get,

\begin{equation}\label{eq12}
\dot{ \mathit{\Lambda}}= \frac{ \alpha }{1- \alpha }~ \left( 8 \pi G\dot{ \rho } \right) 
\end{equation}

Substituting (12) in Energy conservation equation (6),

\begin{equation}\label{eq13}
\dot{ \rho } + 3\frac{\dot{a}}{a} \left(  \rho +P \right) = -\frac{\frac{ \alpha }{1- \alpha }~ \left( 8 \pi G\dot{ \rho } \right) }{8 \pi G} 
\end{equation}

(13) simplifies to,

\begin{equation}\label{eq14}
\dot{ \rho }+ 3\frac{\dot{a}}{a} \rho  \left( 1- \alpha  \right)  \left( 1+ \omega  \right) = 0 
\end{equation}

Replacing the time derivative in (14) with derivative w.r.t scale factor (\textit{a}) and integrating, the expression of density can be readily obtained as,
\begin{equation}\label{eq15}
\rho =Ca^{-3 \left( 1- \alpha  \right)  \left( 1+ \omega  \right) } 
\end{equation} 

(where \textit{C} is the integration constant)

Writing present day density as {\textit{$ \rho$\textsubscript{0}} and noting that present day normalised scale factor is given by (\textit{a\textsubscript{0} }= 1), equation (15) can be written as,

\begin{equation}\label{eq16}
\rho = \rho _{0}a^{-3 \left( 1- \alpha  \right)  \left( 1+ \omega  \right) } 
\end{equation} 

Substituting the ansatz  \(  \mathit{\Lambda} _{v}=3 \alpha  \left( \frac{\dot{a}}{a} \right) ^{2} \) in (8), we obtain the vacuum energy density as,

\begin{equation}\label{eq17}
\rho _{ \mathit{\Lambda}}=\frac{ \mathit{\Lambda} _{C~}+ 3 \alpha  \left( \frac{\dot{a}}{a} \right) ^{2}}{8 \pi G} 
\end{equation} 

Substituting (3) in (17) with \textit{k} = 0,

\begin{equation}\label{eq18}
\rho _{ \mathit{\Lambda}}=\frac{ \mathit{\Lambda} _{C~}+ 3 \alpha  \left[ \frac{8 \pi G \rho }{3}+ \frac{ \mathit{\Lambda}  \left( t \right) }{3} \right] }{8 \pi G} 
\end{equation} 

Equation (18) simplifies to:

\begin{equation}\label{eq19}
\rho _{ \mathit{\Lambda}}=  \frac{\mathit{\Lambda}_C}{8\pi G} +  \alpha  \rho +  \alpha  \rho _{ \mathit{\Lambda}} 
\end{equation} 

Substituting the value of $ \rho $  from (16) in (19) and simplifying, we finally obtain the vacuum energy density as,

\begin{equation}\label{eq20}
\rho _{ \mathit{\Lambda}}= \left( \frac{ \alpha }{1- \alpha } \right)  \rho _{0}a^{-3 \left( 1- \alpha  \right)  \left( 1+ \omega  \right) } + \rho_{\mathit{\Lambda}_{C\alpha}} 
\end{equation} 

where we have defined $\rho_{\mathit{\Lambda}_{C\alpha}} =  \frac{\mathit{\Lambda}_C}{8\pi G(1 - \alpha)}$ as the constant component of vaccum energy density corresponding to this model.

\subsection{Solution for Phenomenological model  \(  \mathit{\Lambda} _{v}=  \beta   \left( \frac{\ddot{a}}{a} \right)  \) }

The ansatz  \(  \mathit{\Lambda} _{v}=  \beta   \left( \frac{\ddot{a}}{a} \right)  \)  gives:

\begin{equation}\label{eq21}
\mathit{\Lambda}  \left( t \right) = \mathit{\Lambda} _{c} + \beta   \left( \frac{\ddot{a}}{a} \right) 
\end{equation}

Substituting (21) in (5) gives,

\begin{equation}\label{eq22}
\frac{ \mathit{\Lambda}  \left( t \right) - \mathit{\Lambda} _{c}}{ \beta }= -\frac{4 \pi G \rho }{3} \left( 1+3 \omega  \right) + \frac{ \mathit{\Lambda}  \left( t \right) }{3}
\end{equation} 

Equation (22) simplifies to,

\begin{equation}\label{eq23}
\mathit{\Lambda}  \left( t \right) =  \left( \frac{ \beta }{ \beta -3} \right) 4 \pi G \rho  \left( 1+3 \omega  \right) +  \left( \frac{3}{3- \beta } \right)  \mathit{\Lambda} _{C} 
\end{equation} 

Taking derivative of (23) w.r.t time, we get,

\begin{equation}\label{eq24}
\dot{ \mathit{\Lambda}}=  \left( \frac{ \beta }{ \beta -3} \right) 4 \pi G\dot{ \rho } \left( 1+3 \omega  \right)  
\end{equation} 

 Substituting (24) in (6),

\begin{equation}\label{eq25}
\dot{ \rho }+3\frac{\dot{a}}{a} \rho  \left( 1+ \omega  \right) = -\frac{ \left( \frac{ \beta }{ \beta -3} \right) 4 \pi G\dot{ \rho } \left( 1+3 \omega  \right) }{8 \pi G}
\end{equation} 

(25) simplifies to,

\begin{equation}\label{eq26}
\dot{ \rho }+3\frac{\dot{a}}{a} \rho  \left( 1+ \omega  \right)  \left[ \frac{2 \beta -6}{3 \beta -6+3 \beta  \omega } \right] =0 
\end{equation}

Replacing time derivative with derivative w.r.t scale factor(\textit{a}) and integrating as before, we get the expression of density in this case as, 

\begin{equation}\label{eq27}
\rho =  \rho _{0}a^{-3 \left( 1+ \omega  \right)  \left( \frac{2 \beta  - 6}{3 \beta  - 6 + 3 \beta  \omega } \right) }
\end{equation}

(where the present-day density  \(  \rho _{0}= C  \)  with \textit{C} being the constant of integration as before)

Vacuum energy density in this case will be given by substituting the ansatz  \(  \mathit{\Lambda} _{v}=  \beta   \left( \frac{\ddot{a}}{a} \right)  \) \  in (8),

\begin{equation}\label{eq28}
 \rho_{ \mathit{\Lambda}}= \frac{\mathit{\Lambda}\left(t\right) }{8 \pi G}= \frac{ \mathit{\Lambda} _{C~}+ \beta   \left( \frac{\ddot{a}}{a} \right) }{8 \pi G}
\end{equation} 

Substituting\ the value of   \( \ddot{\frac{a}{a}} \)  from (4), we get,

 \begin{equation}\label{eq29}
\rho _{ \mathit{\Lambda}}= \frac{ \mathit{\Lambda}_{C~}+ \beta  \left[ -\frac{4 \pi G}{3} \left( 1+3 \omega  \right)  \rho + \frac{ \mathit{\Lambda}  \left( t \right) }{3} \right] }{8 \pi G}
\end{equation}

(29) simplifies to 

\begin{equation}\label{eq30}
\rho _{ \mathit{\Lambda} }=  \frac{\mathit{\Lambda}_C}{8\pi G} - \frac{ \beta  \rho  \left( 1+3 \omega  \right) }{6}+ \frac{ \beta }{3} \rho _{ \mathit{\Lambda} } 
\end{equation}

Substituting (27) in (30), we finally get the vacuum energy density as,

\begin{equation}\label{eq31}
\rho _{ \mathit{\Lambda}}= \rho _{ \mathit{\Lambda} _{C\beta}} + \frac{ \beta  \left( 1+3 \omega  \right) }{2 \left(  \beta -3 \right) } \rho _{0}a^{-3 \left( 1+ \omega  \right)  \left( \frac{2 \beta -6}{3 \beta -6+3 \beta  \omega } \right) }
\end{equation}

where we have defined $\rho_{\mathit{\Lambda}_{C\beta}} = \left( \frac{3}{3- \beta } \right) \frac{\mathit{\Lambda}_C}{8\pi G}$ as the constant component of vacuum energy density corresponding to this model.

\subsection{Solution for Phenomenological model  \(  \mathit{\Lambda} _{v}= 8 \pi G \gamma  \rho  \)}
The ansatz  \(  \mathit{\Lambda} _{v}= 8 \pi G \gamma  \rho  \)  readily gives,

\begin{equation}\label{eq32}
\mathit{\Lambda}  \left( t \right) = \mathit{\Lambda}_{c} +8 \pi G \gamma  \rho 
\end{equation}

Taking time derivative of (32), we get,

\begin{equation}\label{eq33}
\dot{ \mathit{\Lambda}} = 8 \pi G \gamma \dot{ \rho } 
\end{equation}

Substituting (33) in (6) we get,

\begin{equation}\label{eq34}
\dot{ \rho } + 3\frac{\dot{a}}{a} \left(  \rho +P \right) = -\frac{8 \pi G \gamma \dot{ \rho }}{8 \pi G} 
\end{equation}

(34) simplifies to 

\begin{equation}\label{eq35}
\dot{ \rho } \left( 1+ \gamma  \right) + 3\frac{\dot{a}}{a} \rho  \left( 1+ \omega  \right) = 0 
\end{equation}

Replacing time derivative with derivative w.r.t scale factor (\textit{a}) and integrating as previous models, we get the expression of density in this case as,

\begin{equation}\label{eq36}
 \rho = \rho _{0}a^{\frac{-3 \left( 1+ \omega  \right) }{1+ \gamma }} 
\end{equation}

(where the present-day density  \(  \rho _{0}= C  \)  with \textit{C} being the constant of integration as in previous cases)

Vacuum energy density for this model can be obtained by substitution of the ansatz            \(  \mathit{\Lambda} _{v}= 8 \pi G \gamma  \rho  \)  in (8),\par

\begin{equation}\label{eq37}
\rho _{ \mathit{\Lambda}}= \frac{ \mathit{\Lambda}  \left( t \right) }{8 \pi G}= \frac{ \mathit{\Lambda} _{c} +8 \pi G \gamma  \rho }{8 \pi G} 
\end{equation}

(37) simplifies to,

\begin{equation}\label{eq38}
\rho _{ \mathit{\Lambda} }=  \frac{\mathit{\Lambda}_{C}}{8\pi G}+  \gamma  \rho 
\end{equation}

Substituting (36) in (38), we get the expression of vacuum energy density as,

\begin{equation}\label{eq39}
\rho _{ \mathit{\Lambda}}=  \rho _{ \mathit{\Lambda} _{C\gamma}}+  \gamma  \rho _{0}a^{\frac{-3 \left( 1+ \omega  \right) }{1+ \gamma }}
\end{equation}

where we have defined $\rho _{ \mathit{\Lambda} _{C\gamma}} = \frac{\mathit{\Lambda}_{C}}{8\pi G}$ as the constant component of vacuum energy density corresponding to this model.

For all the three types of model explored in this section, the expression of density shows that the dilution of the other major component with the expansion of Universe is dependent upon the variable component of Cosmological Constant  \(  \left(  \mathit{\Lambda} _{v} \right)  \)  since it includes the parameters  \(  \left(  \alpha , \beta , \gamma  \right)  \) . This feature is an indication that  \(  \left(  \mathit{\Lambda} _{v} \right)  \)  decays into the other major component of the Universe in our two-fluid approach\footnote{This physical scenario will only hold if density of the other major component (\textit{$\mathit{\rho}$}) dilutes in our model at a rate slower than \textit{$\mathit{\Lambda}$}CDM model which translates to the constraints $0<$ \textit{$\mathit{\alpha}$} $<1$, \textit{$\mathit{\beta}$} $<0$ and \textit{$\mathit{\gamma}$} $>0$ on the model parameters respectively} which results in modification of the dilution rate of the other major\ component in accordance with the variation of   \(  \mathit{\Lambda} _{v} \). We therefore have a $``$vacuum/dark energy decaying into other major component$"$  picture and for convenience we will call it $``$ decaying vacuum product component$"$. In section 5, we will look into the possible candidates which can be the product component. 

The expressions of density of the vacuum/dark energy points out that in the limit  \( t \rightarrow \infty \)  i.e.  \( a \rightarrow \infty \) ,  \(  \rho _{ \mathit{\Lambda}} \rightarrow \rho_{\mathit{\Lambda}_C} \), which clearly reflects the parallels between the quantum metastable dark energy decay and phenomenological hybrid dynamic \textit{$ \mathit{\Lambda}$} models. Furthermore, the variable component of vacuum and the decaying vacuum product component of the Universe dilutes at the same rate. This feature leads to redefining the $``$coincidence problem$"$  as we shall see later.

\section{EQUIVALENCE OF THE MODELS AND DILUTION RATE PARAMETRIZATION}\label{sec3}

Ray, Mukhopadhyay and Meng~\cite{Ray1} showed that the models  \( \mathit{\Lambda} = 3 \alpha  \left( \frac{\dot{a}}{a} \right) ^{2} \) ,  \(  \mathit{\Lambda} = \beta  \frac{\ddot{a}}{a} \) and \(  \mathit{\Lambda} = 8 \pi G \gamma  \rho  \)  become equivalent when written in terms of dimensionless density parameters. Since from a mathematical point of view, our models are essentially similar to these models with an additive constant, it can be expected that the equivalence should hold for our hybrid models as well. The equivalence can indeed be shown in a straightforward manner by introducing a new parameter \textit{u} in the models which is the exponent of scale factor in the expression of density of the decaying vacuum product component representing how it dilutes with the expansion of the Universe in the presence of dynamic \textit{$\mathit{\Lambda}$.} We will identify this parameter as $``$dilution parameter$"$.

For the three models used in previous section, we can write the corresponding dilution parameters as ,

\begin{equation}\label{eq40}
u_{ \alpha }= 3 \left( 1- \alpha  \right)  \left( 1+ \omega  \right) ~;~~u_{ \beta }=3 \left( 1+ \omega  \right)  \left( \frac{2 \beta -6}{3 \beta -6+3 \beta  \omega } \right) ;~~u_{ \gamma }=\frac{3 \left( 1+ \omega  \right) }{1+ \gamma }
\end{equation}

(where we have adopted the notational convention of adding the model parameters  \(  \alpha , \beta , \gamma  \) as suffixes to the dilution parameter in order to denote the respective models)

For the model  \(  \mathit{\Lambda} _{v}=3 \alpha  \left( \frac{\dot{a}}{a} \right) ^{2} \) the density parameters in equation (16) and (20) of the model can then be written as,

\begin{equation}\label{eq41}
\rho = \rho _{0}a^{-u_{ \alpha }}\text{~ ; } \rho _{ \mathit{\Lambda}}= \left( \frac{3 \left( 1+ \omega  \right) - u_{ \alpha }}{u_{ \alpha }} \right)  \rho _{0}a^{-u_{ \alpha }} +  \rho _{ \mathit{\Lambda}_{C\alpha}}
\end{equation}

Similarly,\ for model   \(  \mathit{\Lambda} _{v}=  \beta   \left( \frac{\ddot{a}}{a} \right)  \) the\ density parameters in (27) and  (31) can be written as,

\begin{equation}\label{eq42}
\rho = \rho _{0}a^{-u_{ \beta }}~;~ \rho _{ \mathit{\Lambda}}= \left( \frac{3 \left( 1+ \omega  \right) - u_{ \beta }}{u_{ \beta }} \right)  \rho _{0}a^{-u_{ \beta }} +  \rho _{ \mathit{\Lambda}_{C\beta}}
\end{equation}

Finally, for the model  \(  \mathit{\Lambda} _{v}= 8 \pi G \gamma  \rho  \) , the density parameters in (36) and (39) will be given by,

\begin{equation}\label{eq43}
\rho = \rho _{0}a^{-u_{ \gamma }}~;~ \rho _{ \mathit{\Lambda}}= \left( \frac{3 \left( 1+ \omega  \right) - u_{ \gamma }}{u_{ \gamma }} \right)  \rho _{0}a^{-u_{ \gamma }} +  \rho _{ \mathit{\Lambda}_{C\gamma}}
\end{equation}

It can be clearly seen from equations (41), (42) and (43), that when expressed in terms of dilution parameter, the evolution of the density parameters in terms of scale factor follows exactly the same pattern for all three models. As a consequence, the three models cannot be distinguished from each other and can be viewed as equivalent. Henceforth, without any loss of generality, we can write,

\begin{subequations}
\begin{equation}\label{eq44a}
u_{ \alpha }= u_{ \beta }= u_{ \gamma }=u 
\end{equation}
\begin{equation}\label{eq44b}
\rho_{\mathit{\Lambda}_{C\alpha}} = \rho_{ \mathit{\Lambda}_{C\beta}} = \rho_{\mathit{\Lambda}_{C\gamma}} = \rho_{ \mathit{\Lambda}_{C}}
\end{equation}
\end{subequations}

The density parameters for all the three models used in this work can then be simply written, without reference to any particular model as,

\begin{equation}\label{eq45}
\rho = \rho _{0}a^{-u}~;~ \rho _{ \mathit{\Lambda} }= \left( \frac{3 \left( 1+ \omega  \right) - u}{u} \right)  \rho _{0}a^{-u} + \rho _{ \mathit{\Lambda}_{C}}
\end{equation}

Here \textit{u} can be simply interpreted as a common $``$dilution parameter$"$  which represents the dilution rate of decaying vacuum product component for all the three models described in this paper and $\rho _{ \mathit{\Lambda}_{C}}$ can be interpreted as the common $``$constant component of vacuum energy density$"$ representing the limiting value of vacuum energy density corresponding to the limit $a \rightarrow \infty$ for all the three models described in this work. From this point onwards we will adopt this common parametrisation for the remaining part of the paper without reference to any particular model. However, if desired one can quickly recover the expressions for any particular model by substituting the respective model parameters in place of \textit{u} by following equation (40).

Equation (44a) and (40) readily gives us an equation connecting the parameters of the three models as,

\begin{equation}\label{eq46}
\left( 1- \alpha  \right) =  \left( \frac{2 \beta -6}{3 \beta -6+3 \beta  \omega } \right) = \frac{1}{1+ \gamma }
\end{equation}

This is exactly the same relation that connected the model parameters in phenomenological models  \(  \mathit{\Lambda} = 3 \alpha  \left( \frac{ȧ}{a} \right) ^{2} \),  \(  \mathit{\Lambda} = \beta  \frac{\ddot{a}}{a} \)  and  \(  \mathit{\Lambda} = 8 \pi G \gamma  \rho  \) ~\cite{Ray1}. Therefore, not only are the hybrid dynamic models equivalent but the relation connecting the model parameters are also exactly same with their pure dynamic model counterparts. Such equivalence among Kinematic models may be attributed to the fact that the underlying phenomenological expressions are introduced from the common ground of dimensional consistency.

\section{COSMOLOGY WITH HYBRID DYNAMIC $ \mathit{\Lambda}$  AND DILUTION PARAMETER}\label{sec4}

It was highlighted in section II, that the background geometry of the Universe will be described by FLRW metric since GR and Cosmological Principle has been assumed to hold in this work. As a result, the hybrid dynamic variation of \textit{$\mathit{\Lambda}$} will have no impact on the useful geometrical relations of standard FLRW Cosmology and we can use them straightaway. The impact of the variation of \textit{$\mathit{\Lambda}$} as opposed to a constant one lies on the Cosmological equations which will be modified. After finding the solutions of Cosmological field equations and developing the general parametrisation in the previous sections, we will now derive some of the modified Cosmological relations.

\subsection{Dimensionless density parameters}
We will adopt the standard definition of dimensionless density parameter  \(  \left(  \mathit{\Omega} _{i}= \frac{8 \pi G}{3H^{2}} \rho _{i} \right)  \)  and express the decaying vacuum product component density as,

\begin{equation}\label{eq47}
\mathit{\Omega} = \frac{8 \pi G}{3H^{2}} \rho = \frac{8 \pi G}{3H^{2}} \rho _{0}a^{-u}
\end{equation}
(where \textit{H} is Hubble parameter)

Similarly, the vacuum density can be expressed as,

\begin{equation}\label{eq48}
\mathit{\Omega} _{ \mathit{\Lambda}}= \frac{8 \pi G}{3H^{2}} \rho _{ \mathit{\Lambda}}= \frac{8 \pi G}{3H^{2}} \left( \frac{3 \left( 1+ \omega  \right) - u}{u} \right)  \rho _{0}a^{-u} + \frac{8 \pi G}{3H^{2}}  \rho _{ \mathit{\Lambda} _{C}}
\end{equation}

Using (47), equation (48) can be written as,

\begin{equation}\label{eq49}
\mathit{\Omega} _{ \mathit{\Lambda}}= \left( \frac{3 \left( 1+ \omega  \right) - u}{u} \right)  \mathit{\Omega} +  \mathit{\Omega} _{ \mathit{\Lambda} _{C}} 
\end{equation}

(where we have defined  \( \mathit{\Omega}_{ \mathit{\Lambda} _{C}}=\frac{8 \pi G}{3H^{2}} \rho_{ \mathit{\Lambda}_{C}} \) representing the dimensionless density parameter corresponding to constant component of vacuum density)

The corresponding present-day dimensionless density parameters can be readily defined from (47), (48) and (49) as,

\begin{subequations}
\begin{equation}
\mathit{\Omega}_{0}= \frac{8 \pi G}{3H_{0}^{2}} \rho _{0}\
\end{equation}
\begin{equation}
\mathit{\Omega}_{ \mathit{\Lambda}_{C0}}= \frac{8 \pi G}{3H_{0}^{2}} \rho _{ \mathit{\Lambda}_{C0}} \
\end{equation}
\begin{equation}
\mathit{\Omega}_{\mathit{\Lambda}_{0}}= \frac{8 \pi G}{3H_{0}^{2}} \rho _{ \mathit{\Lambda}_{0}}=  \left( \frac{3 \left( 1+ \omega  \right) - u}{u} \right)  \mathit{\Omega}_{0} +  \mathit{\Omega}_{\mathit{\Lambda}_{C0}}
\end{equation}
\end{subequations}

(where we have used the notational convention of adding ‘zero’ as suffix to denote present-day values of parameters)

Using (47) and (48), and definition of Hubble parameter  \(  \left( H= \frac{\dot{a}}{a} \right)  \)  we can write the Cosmological field equation (3) in terms of dimensionless density parameters as,

 \begin{equation}\label{eq51}
\mathit{\Omega} +  \mathit{\Omega}_{ \mathit{\Lambda}}=1 
\end{equation}

(where we have assumed Flat Universe with  \( k = 0 \) )

Substituting (49) in (51) we get,

\begin{equation}\label{eq52}
 \mathit{\Omega} \left[\frac{3(1 + \omega)}{u}\right] +  \mathit{\Omega}_{ \mathit{\Lambda}_{C}} = 1 
\end{equation}

Equation (52) is the field equation for our phenomenological hybrid dynamic \textit{$ \mathit{\Lambda}$} model in terms of dimensionless density parameters which must hold at all epoch for a flat Universe. This equation is readily used in Computational Cosmology to extract the model parameters for a flat Universe.

\subsection{Hubble parameter equation}
A very important parameter in Cosmology is the Hubble parameter \textit{(H)} which measures the expansion rate of the Universe. Mathematically, it is defined in terms of normalised scale factor as   \( H=\frac{\dot{a}}{a} \). The Cosmological field equation (3) when written in terms of present-day dimensionless density parameters, using (50) gives the Hubble parameter expression,

\begin{equation}\label{eq53}
H^{2}= H_{0}^{2} \left[ \frac{3 \left( 1+ \omega  \right) }{u} \mathit{\Omega}_{0}a^{-u} + \mathit{\Omega}_{ \mathit{\Lambda}_{C0}} \right] 
\end{equation}

(where  \( H_{0} \) \ denotes the present-day Hubble parameter  i.e. Hubble constant)

A useful relation that comes from FLRW geometry is the relation  \( a= \left( 1+z \right) ^{-1} \)  connecting redshift parameter \textit{(z)} and normalised scale factor \textit{(a)}. Using it, the Hubble parameter equation (53) can be written in terms of redshift as,

\begin{equation}\label{eq54}
{H\left(z\right)}^2=\ H^2_0\left[\frac{3\left(1+\omega \right)}{u}{\mathit{\Omega}}_0{\left(1+z\right)}^u+{\mathit{\Omega}}_{{\mathit{\Lambda}}_{C0}}\right] 
\end{equation}

The Hubble parameter equation (53, 54) is perhaps the backbone of Cosmology since many other useful relations are derived with it. It builds up into $``$Luminosity distance$"$  and $``$Angular Diameter distance$"$  equations which are vividly used in Observational Cosmology and as well into the expression of look-back time which is another vital relation of Cosmology.

\subsection{Deceleration parameter and transition redshift}

The acceleration or deceleration of the expansion rate of the Universe is characterized in Cosmology by the dimensionless deceleration parameter \textit{q}\ which is mathematically defined as   \(  \left( q=- \frac{1}{H^{2}}\frac{\ddot{a}}{a} \right)  \) . An expression for deceleration parameter can be obtained by writing the Cosmological field equation (5) in terms of dimensionless density parameters using (47-49) which gives,

\begin{equation}\label{eq55}
q=\mathit{\Omega}\left[\frac{\left(1+\omega \right)\left(3u - 6\right)}{2u}\right] - \mathit{\Omega}_{\mathit{\Lambda_C}}
\end{equation}

Equation (55) is the deceleration parameter equation for our model. It can be expressed in a convenient form by writing (55) in terms of present-day density parameters using (47-50) and redshift parameter\textit{(z)} using  \( a= \left( 1+z \right) ^{-1} \)  which gives ,

\begin{equation}\label{eq56}
q= \frac{H_{0}^{2}}{H^{2}} \left[\mathit{\Omega_0}\left(1+z\right)^u\left(\frac{\left(1+\omega \right)\left(3u - 6\right)}{2u}\right) - \mathit{\Omega}_{\mathit{\Lambda_{C0}}}\right] 
\end{equation}

For the present Universe to expand at an accelerating rate, the present-day deceleration parameter must have negative sign. However, structure formation in the Universe cannot take place during acceleration and the accelerating phase must be preceded by a decelerating phase ~\cite{Ray2}. Observational studies ~\cite{Hinshaw, Aghanim} also support the idea that Cosmic acceleration is a recent phenomenon. This means that at some epoch the deceleration parameter must undergo a change of sign in any physically sensible Cosmological model. At this juncture, we will highlight an important feature of the hybrid dynamic nature of \textit{$\mathit{\Lambda}$}  that we adopted in this work. The additive constant (\(  \mathit{\Lambda}_{C} \)) in the expression of  \textit{$\mathit{\Lambda}$}  i.e. the constant component of  \textit{$\mathit{\Lambda}$}, manifests itself through the second term in the expression of \textit{q}. The presence of the second term in equation (55) or (56) ensures that the signature flip of \textit{q} can always be achieved in a straight-forward manner. In case of pure dynamic  \textit{$\mathit{\Lambda}$} models, the second term will be absent and as a consequence, the characteristic sign change of \textit{q} cannot be readily achieved with two-fluid assumption and such model Universes will either be always accelerating or always decelerating which is physically absurd.

The transition point of the Universe from a decelerating phase to an accelerating phase is characterized by null value of the deceleration parameter. The corresponding redshift is the transition redshift  \(  \left( z_{t} \right)  \) . An expression for transition redshift can be obtained by substituting (52) and (54) in (56) and setting  \( q = 0 \)  to represent the transition point. It readily gives, 

\begin{equation}\label{eq57}
z_{t}= \left[\frac{2u - 6\mathit{\Omega_0}\left(1 + \omega\right)}{\left(1 + \omega\right)\left(3u - 6\right)\mathit{\Omega_0}}\right]^{\frac{1}{u}}-1
\end{equation}

In section VII, we will estimate the value of present-day deceleration parameter and transition redshift for our model.

\subsection{Distance relations: Luminosity distance and Angular diameter distance}

In FLRW Cosmology, the proper distances to an object cannot be measured and as such $``$distances$"$  are usually specified by $``$Luminosity distance$"$  and $``$Angular diameter distance$"$  which simply corresponds to measuring distances in terms of luminosity and angular diameters of astrophysical objects respectively. These quantities are widely used for observational studies and therefore it is necessary to find the modified expressions of these quantities for our model. 

In FLRW geometry, Luminosity distance (\( d_{L} \)) and Angular diameter distance (\( d_{A} \)) for flat Universe (\( k = 0 \)) are expressed as,

\begin{equation}\label{eq58}
d_{L}= \chi  \left( 1+z \right) ~;~~~~~~~~~~~~~~d_{A}=\frac{ \chi }{ \left( 1+z \right) }
\end{equation}

(Where  \(  \chi  \)  is co-moving coordinate of the astrophysical source, which is defined in terms of Hubble parameter as  \(  \chi = \int _{0}^{z}\frac{dz}{H \left( z \right) } \)  )

Using the Hubble parameter expression (54), we can write the luminosity distance expressions as,

\begin{equation}\label{eq59}
d_L=\ \frac{\left(1+z\right)}{H_0}\int^z_0{\frac{dz}{\sqrt{\frac{3\left(1+\omega \right)}{u}{\mathit{\Omega}}_0{\left(1+z\right)}^u+{\mathit{\Omega}}_{{\mathit{\Lambda}}_{C0}}}}}
\end{equation}

Similarly, using (54), the angular diameter distance can be expressed as,

\begin{equation}\label{eq60}
d_A=\ \frac{1}{{\left(1+z\right)H}_0}\int^z_0{\frac{dz}{\sqrt{\frac{3\left(1+\omega \right)}{u}{\mathit{\Omega}}_0{\left(1+z\right)}^u+{\mathit{\Omega}}_{{\mathit{\Lambda}}_{C0}}}}}
\end{equation}

In particular we will be using the luminosity distance relation (59) to confront our model with observations and extract the values of Cosmological parameters by using type 1a supernova method. This will be explained in more details in section VI.

\subsection{Look-back time and Age of the Universe}

Another important expression in analysis of Cosmological models is the $``$look-back time$"$  which refers to the difference between the cosmic time in which a galaxy emitted a photon (\textit{t}) and present cosmic time when it is received by us (\textit{t\textsubscript{0}}). From FLRW geometry, the look-back time is given in terms of redshift as,

\begin{equation}\label{eq61}
t_{0}-t= \int _{0}^{z}\frac{dz}{ \left( 1+z \right) H \left( z \right) } 
\end{equation}

 Using (54), the expression of look-back time for our model can be readily obtained as,

\begin{equation}\label{eq62}
t_0-t=\frac{1}{H_0}\int^z_0{\frac{dz}{\left(1+z\right)\sqrt{\left[\frac{3\left(1+\omega \right)}{u}{\mathit{\Omega}}_0{\left(1+z\right)}^u+{\mathit{\Omega}}_{{\mathit{\Lambda}}_{C0}}\right]}}}
\end{equation}

In Cosmological models which has a Big-Bang origin, a very important quantity is the Cosmic Age or Age of the Universe which refers to the time that has elapsed between the epoch where scale factor  \( a \left( t \right)  = 0 \)  and present epoch (\textit{t\textsubscript{0}}). The expression of Cosmic Age can be easily obtained from the expression of look-back time (62) by putting (\( z \rightarrow \infty \)) which gives,

\begin{equation}\label{eq63}
t_0=\frac{1}{H_0}\int^{\infty }_0{\frac{dz}{\left(1+z\right)\sqrt{\left[\frac{3\left(1+\omega \right)}{u}{\mathit{\Omega}}_0{\left(1+z\right)}^u+{\mathit{\Omega}}_{{\mathit{\Lambda}}_{C0}}\right]}}}
\end{equation}
Analysis\ of Cosmic Age is a very important aspect of Cosmological models because of its connection with the historic Cosmic Age Problem which refers to the puzzling situation of finding age of the Universe in a Cosmological model to be less than ages of some objects in the Universe. We will come back to this topic in details in section VII where we will estimate Cosmic Age for our model.

\subsection{Particle number density and creation rate}
In section II, it was seen that the solutions of the three models hinted towards a physical scenario of vacuum decaying into the other major component. Naturally, this implies creation of particles of the decaying vacuum product component\footnote{In Appendix A, we have explored an alternative scenario where instead of creation of particles, decay of vacuum is causing an increase in mass of the particles of decaying vacuum product component} from decay of vacuum. If \textit{n} denotes the number density and \textit{m} denotes the mass of the product component, then assuming particles are mass invariant, the conservation equation (6) can be written as, 

\begin{equation}\label{eq64}
m\dot{n}\ +\ 3\frac{\dot{a}}{a}mn \left( 1 + \omega \right) = - \dot{\rho_{\mathit{\Lambda}}}
\end{equation}

Equation (64) can be conveniently expressed in the form,

\begin{equation}\label{eq65}
\dot{n}\ +\ 3\frac{\dot{a}}{a}n \left( 1 + \omega \right) = - n\left(\frac{\dot{\rho_{\mathit{\Lambda}}}}{\rho}\right)
\end{equation}

In Cosmological models involving particle creation or annihilation in an FLRW background, a general equation for the number density of the relevant component of cosmological fluid can be obtained~\cite{Lima2} as,

\begin{equation}\label{eq66}
\dot{n} + 3\frac{\dot{a}}{a}n \left( 1 + \omega \right) = \psi = n\Gamma
\end{equation}

In equation (66) $\psi>0$ for particle sources and $\psi < 0$ for particle sinks.  Particle creation or annihilation rate parameter defined by $\Gamma = \psi/n$ in the general equation can be related to any physical process which creates or annihilates particles of the relevant component of Cosmological fluid. In the context of our model, $\Gamma$ will relate to decay of vaccum energy and corresponding creation of particles of the other major component. From equation (65) and (66), we can write the mathematical form of creation rate parameter for our model as,

\begin{equation}\label{eq67}
\Gamma = - \frac{\dot{\rho_{\mathit{\Lambda}}}}{\rho}
\end{equation}  

Using expressions of $\rho$ and $\rho_{\mathit{\Lambda}}$ from equation (45), the expression of particle creation rate parameter becomes,

\begin{equation}\label{eq68}
\Gamma = \left[3\left(1+ \omega \right) - u\right] \frac{\dot{a}}{a}
\end{equation}

Using the expression of $\Gamma$ obtained in expression (68), conservation equation (65) can be written as,

\begin{equation}\label{eq69}
\dot{n} + nu\frac{\dot{a}}{a} = 0
\end{equation}

The above expression can be easily solved to yield,  

\begin{equation}\label{eq70}
n = n_0 a^{-u}
\end{equation}
where $n_0$ denotes the present day value of number density. 
Equation (70) gives the evolution of particle number density in terms of scale factor in the presence of variable Cosmological constant which clearly differes from \textit{$\mathit{\Lambda}$}CDM model where evolution of particle density follows the standard expression $n = n_0 a^{-3(1+\omega)}$.

Particle creation rate parameter can also be written in terms of density parameters and by substituting Hubble parameter equation (53, 54) in (68) which readily gives the following equations for the parameter in terms of scale factor and redshift respectively,

\begin{equation}\label{71}
\Gamma (a) = H_{0} \left[3\left(1+ \omega \right) - u\right]  \left[\sqrt{ \frac{3 \left( 1+ \omega  \right) }{u} \mathit{\Omega}_{0}a^{-u}+ \mathit{\Omega}_{ \mathit{\Lambda}_{C0}}}\right] 
\end{equation}

\begin{equation}\label{72}
\Gamma (z) = H_{0} \left[3\left(1+ \omega \right) - u\right]  \left[\sqrt{ \frac{3 \left( 1+ \omega  \right) }{u}\mathit{\Omega}_{0}(1+ z)^{u}+ \mathit{\Omega}_{ \mathit{\Lambda}_{C0}}}\right] 
\end{equation}

Particle creation rate parameter is an extremely crucial parameter in Cosmologies involving some form of particle creation. In section VII, we will estimate the value of the parameter for our model.

\section{COMPONENTS OF COSMOLOGICAL FLUID}\label{sec5}

So far, we have held the notion that our model Universe is made up of two fluids - a Dark Energy component represented by phenomenological hybrid dynamic Cosmological Constant with E.O.S.  \(  \omega _{ \mathit{\Lambda}}= -1  \)  and a decaying vacuum product component with E.O.S.  \(  \omega  \) . We have built up the model and obtained the relevant Cosmological equations with this consideration. Therefore, in this analysis we are restricted to a two component Cosmological fluid, one of which is Dark Energy. However, the decaying vacuum product component has not been specified yet. In this section we will look at some of the choices for it and will also adopt a particular case for our observational analysis.

\subsection{Choices for the decaying vacuum product component}

Since the E.O.S. parameter \textit{$\omega$} depends solely on the nature of the decaying vacuum product fluid, a choice for the $``$decaying vacuum product component$"$  boils down to the choice for E.O.S parameter. It was highlighted in section 2 that our variable-\textit{$ \mathit{\Lambda}$} approach coincides with the physical scenario of $``$vacuum decaying into the decaying vacuum product component$"$  which naturally sets up a constraint that the physical decay process must be realizable for the model to be realistic. The two most natural candidates for decaying vacuum are pressureless dust with  \(  \omega =0 \) and radiation with   \(  \omega =1/3 \). Cosmology with vacuum decaying into massless radiation has been explored by Freese et al. ~\cite{Freese}. Vacuum decaying into radiation will lead to creation of photons which will have observable impact on microwave background~\cite{Overduin2, Peebles}. Although such a possibility cannot be ruled out completely, observational studies ~\cite{Opher} suggest that even if such a decay process happens, it will be so small that it will be practically equivalent to zero vacuum decay. The other choice which involves Vacuum decaying into dust is feasible. In the introduction section, we highlighted that one of the advantages of using hybrid dynamic nature of \textit{$ \mathit{\Lambda}$} instead of a pure dynamic \textit{$\mathit{\Lambda}$} is that in our approach parallels can be drawn between standard \textit{$\mathit{\Lambda}$}CDM model with constant \textit{$ \mathit{\Lambda}$} and variable-\textit{$ \mathit{\Lambda}$} scenario. Keeping the same spirit, we will adopt pressureless Dust to be the decaying vacuum product component for our observational studies since it reciprocates the standard \textit{$ \mathit{\Lambda}$}CDM scenario with radiation ignored and Dark Energy represented by \textit{$\mathit{\Lambda}$}. The only difference with the standard model being the variation of \textit{$\mathit{\Lambda}$}. For the sake of completeness, we should mention that vacuum decaying into baryonic matter is problematic on grounds of Baryon number conservation ~\cite{Overduin2} and for a realistic case, we should have vacuum decaying into Dark Matter only. However, such a scenario is difficult to handle in our approach since both CDM and baryonic matter correspond to  \(  \omega  = 0 \). Henceforth, we will not distinguish between the two types of matter and instead club them together as $``$Dust$"$. If desired, one can ofcourse assume the decaying vacuum product component to be Dark Matter\ in our parametrization and ignore the presence of Baryons which will replicate the $``$vacuum decaying into dark matter$"$  scenario approximately. With such choice of decaying vacuum product, our model overlaps to some extent with the “vacuum decaying to cold dark matter” scenario explored in ~\cite{Wang1}. However, our approaches are different. While ~\cite{Wang1} postulates a modified expression for cold dark matter density different from the standard case as a consequence of decay of vacuum energy, a priori, our approach on the other hand starts of with phenomenological expressions of  \textit{$ \mathit{\Lambda}$}(\textit{t}) and obtains the density expressions. Furthermore, even though we are eventually using Dust as the decaying vacuum product component, the general theoretical framework of our model do not fix the product component beforehand which naturally opens up the window for further explorations with various choices of the component, including non-conventional ones such as stiff fluid with \(  \omega =2 \) which sterns out from the possibility that Universe might had a stiff fluid era ~\cite{Chavanis}. Finally, we should mention that Basilakos ~\cite{Basilakos} explored a \textit{$ \mathit{\Lambda}$}(\textit{t}) model where the decaying vacuum product component was identified as the dominating component of the Universe i.e. radiation in the radiation era and matter in matter era. The model is interesting and implies that vacuum can decay both into matter and radiation and depending on the era one of the decay process will take prominence. This type of scenario is achievable within the mathematical framework of our model as well but as already noted above decay of vacuum into radiation is unlikely. Henceforth we will stick with "vacuum decaying into dust" scenario for our observational analysis and from this point onwards, the mention of hybrid dynamic \textit{$ \mathit{\Lambda}$} model in this paper will indicate a Universe made up of Dust and Dark Energy (decaying) unless specified otherwise.

\subsection{Cosmological equations for Dust-Dark Energy (decaying) type hybrid dynamic \textit{$ \mathit{\Lambda}$} Universe}

Following up from the discussion in previous sub-section, we will write down the Cosmological equations derived in section 4 for Dust-Dark Energy (decaying) Universe by setting E.O.S. parameter $ \omega $  = 0 in all the equations of section 4 which gives,

\begin{equation}\label{eq73}
\mathit{\Omega}_{m}\left(\frac{3}{u}\right) +  \mathit{\Omega}_{ \mathit{\Lambda}_{C}} = 1
\end{equation}

\begin{subequations}
\begin{equation}\label{eq74a}
H^2 (a) = H^2_0\left[\frac{3}{u}{\mathit{\Omega}}_{m0}a^{-u}+{\mathit{\Omega}}_{{\mathit{\Lambda}}_{C0}}\right]
\end{equation}
\begin{equation}\label{eq74b}
H^2 (z)=\ H^2_0\left[\frac{3}{u}{\mathit{\Omega}}_{m0}(1+z)^{u}+{\mathit{\Omega}}_{{\mathit{\Lambda}}_{C0}}\right]
\end{equation}
\end{subequations}

\begin{subequations}
\begin{equation}\label{eq75a}
q= \mathit{\Omega}_{m}\left(\frac{3u - 6}{2u}\right) - \mathit{\Omega}_{\mathit{\Lambda_C}}
\end{equation}
\begin{equation}\label{eq75b}
q= \frac{H_{0}^{2}}{H^{2}} \left[\mathit{\Omega_{m0}}\left(1+z\right)^u\left(\frac{3u - 6}{2u}\right) - \mathit{\Omega}_{\mathit{\Lambda_{C0}}}\right] 
\end{equation}
\end{subequations}

\begin{equation}\label{eq76}
z_{t} = \left[\frac{2u - 6\mathit{\Omega_{m0}}}{\left(3u - 6\right)\mathit{\Omega_{m0}}}\right]^{\frac{1}{u}}-1
\end{equation}

\begin{equation}\label{eq77}
d_L=\ \frac{\left(1+z\right)}{H_0}\int^z_0{\frac{dz}{\sqrt{\frac{3}{u}{\mathit{\Omega}}_{m0}{\left(1+z\right)}^u+{\mathit{\Omega}}_{{\mathit{\Lambda}}_{C0}}}}}
\end{equation}

\begin{equation}\label{eq78}
d_A=\ \frac{1}{{\left(1+z\right)H}_0}\int^z_0{\frac{dz}{\sqrt{\frac{3}{u}{\mathit{\Omega}}_{m0}{\left(1+z\right)}^u+{\mathit{\Omega}}_{{\mathit{\Lambda}}_{C0}}}}}
\end{equation}

\begin{equation}\label{eq79}
t_0-t=\frac{1}{H_0}\int^z_0{\frac{dz}{\left(1+z\right)\sqrt{\left[\frac{3}{u}{\mathit{\Omega}}_{m0}{\left(1+z\right)}^u+{\mathit{\Omega}}_{{\mathit{\Lambda}}_{C0}}\right]}}}
\end{equation}

\begin{equation}\label{eq80}
t_0=\frac{1}{H_0}\int^{\infty }_0{\frac{dz}{\left(1+z\right)\sqrt{\left[\frac{3}{u}{\mathit{\Omega}}_{m0}{\left(1+z\right)}^u+{\mathit{\Omega}}_{{\mathit{\Lambda}}_{C0}}\right]}}}	
\end{equation}

\begin{equation}\label{eq81}
n_m = n_{m0} a^{-u}
\end{equation}

\begin{subequations}
\begin{equation}\label{eq82a}
\Gamma_{m} (a) = H_{0} \left[3 - u\right]  \left[\sqrt{ \frac{3 }{u}\mathit{\Omega}_{m0}a^{-u}+ \mathit{\Omega}_{ \mathit{\Lambda}_{C0}}}\right]
\end{equation}
\begin{equation}\label{eq82b}
\Gamma_{m} (z) = H_{0} \left[3 - u\right]  \left[\sqrt{ \frac{3}{u}\mathit{\Omega}_{m0}(1+z)^{u}+ \mathit{\Omega}_{ \mathit{\Lambda}_{C0}}}\right] 
\end{equation}
\end{subequations}
(where we have added \textit{m} to the suffix to denote that the other major component is pressureless dust i.e. matter).

\section{CONFRONTING THE MODEL WITH OBSERVATIONS AND ESTIMATING THE PARAMETERS}\label{sec6}

Estimation of model parameters in Cosmology is done using different methods such as type 1a supernova, Baryon Acoustic Oscillations(BAO), Weak Lensing (WL), Galaxy Clusters and Redshift Space distortions (RSD). A combination of different methods is often used in different studies. In this work, however, we have only used the type 1a supernova technique which considers the supernovae to be $``$standard candles$"$  with fixed intrinsic luminosity and estimates the model parameters using Luminosity Distance relation (77). If \textit{M} be the absolute magnitude and \textit{m} be the apparent magnitude of a type 1a supernova, then, the theoretical relation between the luminosity distance and magnitude is given by,

\begin{equation}\label{eq83}
M=m-5\log _{10} \left( \frac{d_{L}}{\text{1 Mpc}} \right) -25 
\end{equation}

Defining distance modulus   \(  \left(  \mu =m-M \right)  \) , equation (83) can be written as, 

\begin{equation}\label{eq84}
\mu _{model}=5\log _{10} \left( \frac{d_{L}}{\text{1 Mpc}} \right) +25 
\end{equation} 

(where the expression of luminosity distance depends on the Cosmological model chosen for investigation thereby making  \(  \mu  \)  model dependent as well which is indicated by suffix)

\subsection{Statistical Procedure}

The measured apparent magnitude  \(  \left( m_{obs} \right)  \)  of a supernova cannot be used to obtain observational value of distance modulus  \(  \left(  \mu _{obs} \right)  \)  directly, rather it has to be corrected for stretch factor, colour, corrections from distance biases etc which are associated with their respective nuisance parameters. The absolute magnitude \textit{M} is also treated as a nuisance parameter. The supernova data from union 2.1 compilation ~\cite{dataset} comprises of a dataset of 580 supernova presented as   \textit{\(  \left(  \mu _{obs}\text{~,  z} \right)  \)}  pairs. The nuisance parameters are set to their global derived values and this data can be readily used for Cosmological model fitting to determine the Cosmological parameters. We will use this sample\footnote{in appendix B, we have included the results of fitting the model with Supernova data from Pantheon sample} for fitting in our model. However, we will fit for matter density (\(  \mathit{\Omega}_{m,0} \)) and dilution parameter (\textit{u}). Hubble constant  \( H_{0} \)  cannot be fitted from supernova data alone since it is degenerate with the value of absolute magnitude \textit{M} ~\cite{Dolgov}. The union 2.1 dataset derived the value of nuisance parameter \textit{M} by setting the value of reduced Hubble constant  \(  \left( \frac{H_{0}}{100}=h=0.7 \right)  \) .\ Therefore, we will also set   \( H_{0}=70 \)  in our fitting procedure since we will use the dataset as it is given in ~\cite{dataset}.

The goodness-of-fit parameter for fitting procedure is given by,

\begin{equation}\label{eq85}
\chi ^{2}= \sum _{i}^{}\frac{ \left(  \mu _{obs_{i}}- \mu _{model} \right) ^{2}}{ \sigma _{i}^{2}}~ 
\end{equation}

The likelihood probability  \textit{(P)} of Cosmological parameters can then be written as,

\begin{equation}\label{eq86}
P \propto -\frac{ \chi ^{2}}{2}
\end{equation}

Here  \(  \mu _{obs_{i}} \)  and  \(  \sigma _{i} \)  represents the observational value and uncertainty corresponding to the redshift  \(  \left( z_{i} \right)  \)  whereas  \(  \mu _{model} \)  is the model-dependent theoretical value of distance modulus obtained from (84). For the fitting procedure, we have extracted the Cosmological parameters by fitting using emcee ~\cite{Mackey1} which is a python module that implements the Affine Invariant Markov Chain Monte Carlo (MCMC) method to estimate the model parameters. In order to carry out the fitting, we built our own personalized fitting code ~\cite{Aich}\footnote{all codes used in this work for statistical analysis, generating plots and calculations can be found in ~\cite{Aich}} using Lmfit module ~\cite{Newville} of python. As a cross-check for the accuracy of the parameter estimates for our fitting procedure, we checked it by fitting it to the base \textit{$ \mathit{\Lambda}$}CDM model (radiation ignored $\&$  Dark Energy represented by constant \textit{$ \mathit{\Lambda}$}) using the same Union 2.1 dataset ~\cite{dataset}. It yielded results  \(  \mathit{\Omega}_{ \mathit{\Lambda} 0}= 0.72  \pm 0.01 \)  for standard \textit{$ \mathit{\Lambda}$}CDM model which is excellent agreement with official release ~\cite{Suzuki} of Union 2.1 project. Therefore, we conclude that our fitting method does produce reasonable estimates of fit parameters and can be used for parameter estimation. 

\subsection{Parameter estimates}

Using the statistical procedure described above, the best fit values for the model parameters for our model have been found to be:

\vspace{\baselineskip}

\[\mathit{\Omega}_{m0} = 0.29 \pm 0.03 ~~~~~~     ; ~~~~~~~    u = 2.90 \pm 0.54 \]
(errors reported are 1$ \sigma$). 

\vspace{\baselineskip}

The Hubble diagram and Corner plots ~\cite{Mackey2} for the fitted parameters for hybrid dynamic \textit{$\mathit{\Lambda}$} model using Union 2.1 dataset is shown in Figure~\ref{fig:Hubble diagram} and Figure~\ref{fig:Corner Plots}, respectively.  In table~\ref{table:fit results}, results of fitting the dataset against different Cosmological models has been shown for comparison.

In our model Universe since Dark Energy is decaying into Dust, Dust must decay slowly compared to standard \textit{$\mathit{\Lambda}$}CDM model which is reflected by the marginal deviation of best-fit values of the parameters compared to the \textit{$\mathit{\Lambda}$}CDM counterparts. The uncertainty in dilution parameter is however quite high which will prevent us to conclude whether our model is better than the standard model or not. A joint analysis combining other methods might be able to bring down the uncertainty. However, as of now we can study the physical features of the Universe using the best-fit values. 

%%%%%%%%%%%%%%%%%%%%%%%%%%%%%%%%%%%

\begin{figure*}

\centering

{\includegraphics[width=5.98 in, keepaspectratio=true]{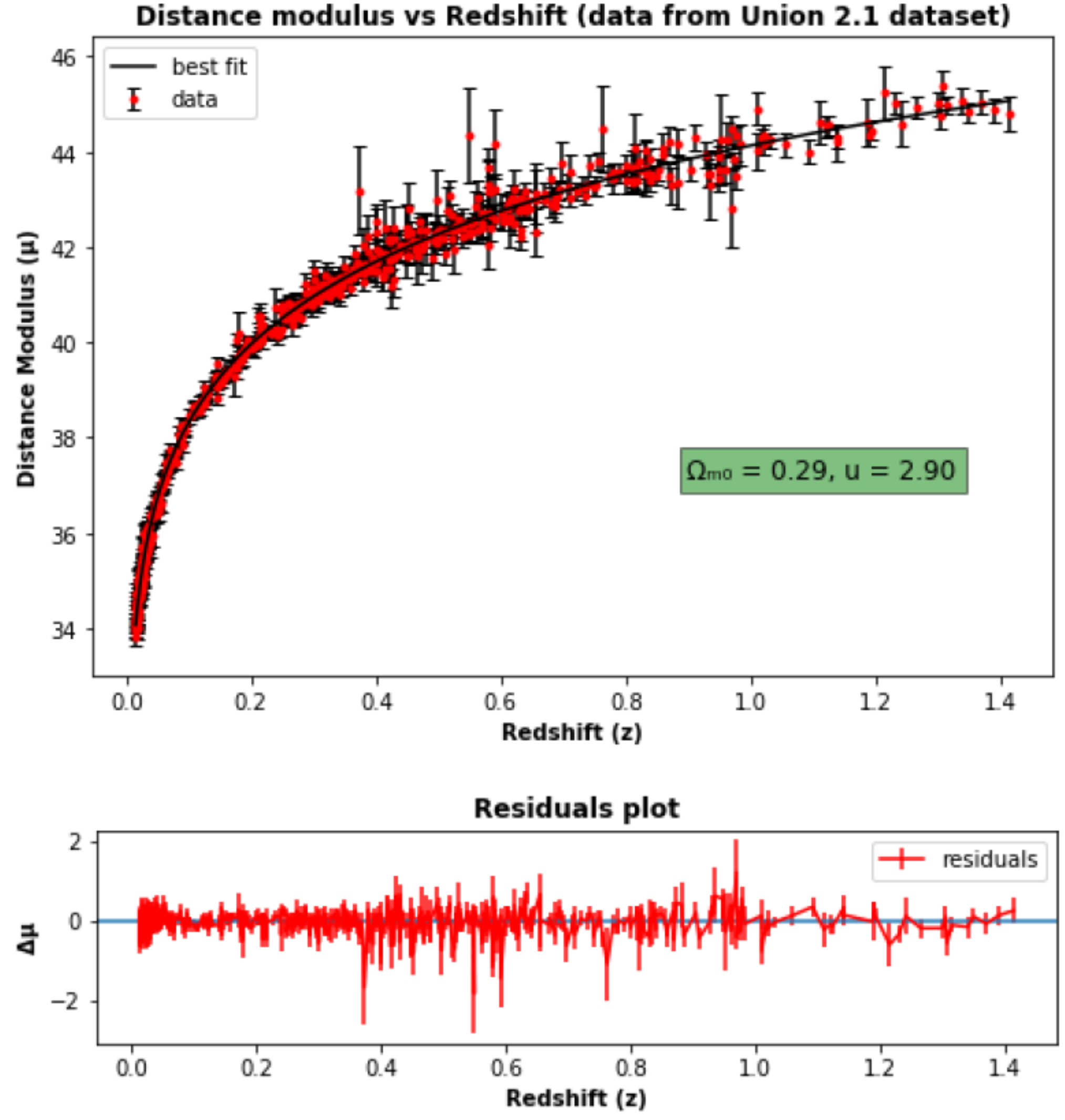}}

\caption{\textbf{\textit{The top plot shows the Hubble diagram using 590 supernova sample of Union 2.1 dataset for hybrid dynamic $\boldsymbol{\mathit{\Lambda}}$ Cosmological model with Dust-Dark Energy (decaying) Universe. The bottom plot shows the corresponding residuals.}}}
\label{fig:Hubble diagram}

\end{figure*}

%%%%%%%%%%%%%%%%%%%%%%%%%%%%%%%%%%%

%%%%%%%%%%%%%%%%%%%%%%%%%%%%%%%%%%%
\begin{figure*}
\centering

\subfigure[] {\includegraphics[width=4.38in, keepaspectratio=true]{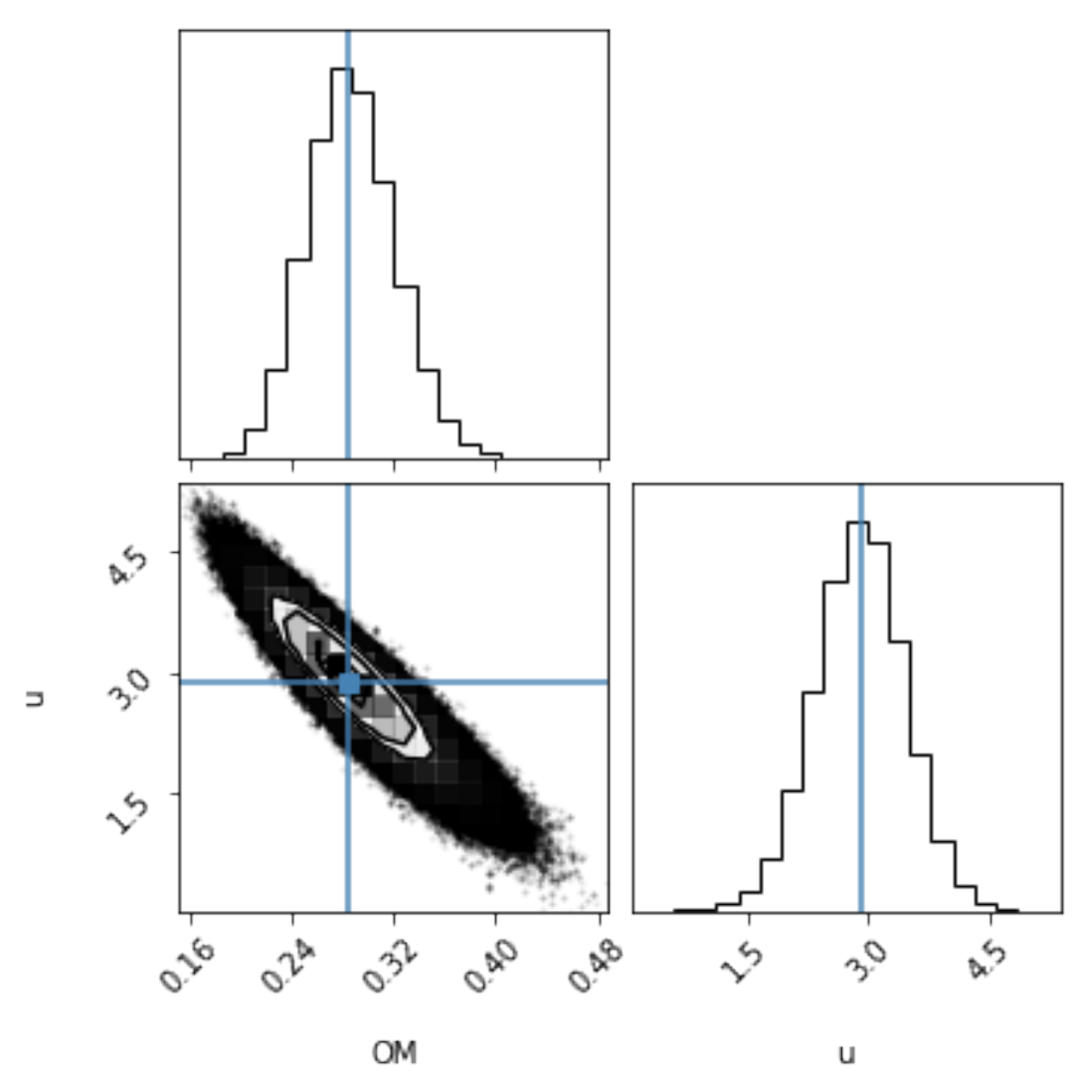}}

\subfigure[] {\includegraphics[width=3.74in, keepaspectratio=true]{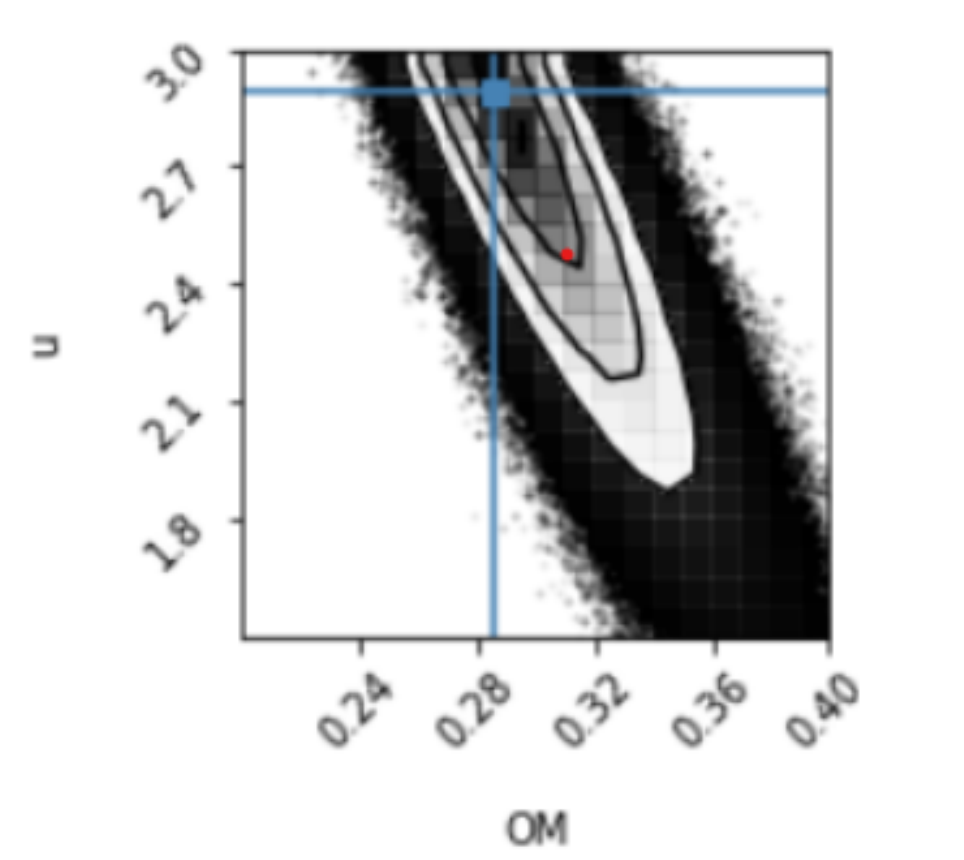}}

\caption{\textbf{\textit{The top figure (a) is the corner plot output showing results of our markov chain monte carlo parameter estimation for hybrid dynamic $\boldsymbol{\mathit{\Lambda}}$ Cosmological model with Dust-Dark Energy (decaying) Universe. The bottom figure (b) is the zoomed in image of the parameter contours which has been terminated at u = 3 to represent realistic scenario. The mark in the bottom figure represents the point ($\boldsymbol{\mathit{\Omega}_{m0}}$ = 0.31,  u = 2.475) which will be discussed later. Note - The label ``OM'' in the corner plots is equivalent to  $\boldsymbol{\mathit{\Omega}_{m0}}$.}}}
\label{fig:Corner Plots}

\end{figure*}

%%%%%%%%%%%%%%%%%%%%%%%%%%%%%%%%%%%

%%%%%%%%%%%%%%%%%%%%%%%%%%%%%%%%%%%

\begin{table}
\centering
\captionsetup{justification=centering}
\caption{Comparison table of fitting different models}

\begin{tabular}{| c |c |c |} \hline 
Model & parameter fit results $(h = 0.7)$ & reduced chi-square \\ \hline
$\mathit{\Lambda}$CDM model  &  $\mathit{\Omega_{\Lambda0}}$ = 0.72 $\pm$ 0.01 & 0.9711 \\ \hline 
o$\mathit{\Lambda}$CDM model & $\mathit{\Omega}_{m0} = 0.29 \pm 0.16, \mathit{\Omega}_{k0} = -0.01 \pm 0.28$ & 0.9729\\ \hline
wCDM model & $\mathit{\Omega_{mo}}$ = 0.28 $\pm$ 0.06, $w$ = -1.00 $\pm$ 0.15 & 0.9727 \\ \hline
\makecell{Hybrid Dynamic \textit{$\mathit{\Lambda}$} model} & ${\mathit{\Omega}}_{m0}$ = 0.29 $\pm$ 0.03, $u$ = 2.90 $\pm$ 0.54 & 0.9728 \\ \hline

\end{tabular}
\label{table:fit results}
\end{table}

\section{COSMOLOGICAL IMPLICATIONS}\label{sec7}

In the last section, we estimated the best fit parameters for our model Universe. Here we will focus on the Cosmological implications of these estimates and look into the physical features of the model.

\subsection{Scale factor evolution}

Using the definition of Hubble parameter, the Hubble parameter equation for Dust-Dark Energy (decaying) Universe (74a) can be written in terms of scale factor as,

\begin{equation}\label{eq87}
{\dot{a}}^2=\ H^2_0\left[\frac{3}{u}{\mathit{\Omega}}_{m0}a^{2-u}+{\mathit{\Omega}}_{{\mathit{\Lambda}}_{C0}}a^2\right]
\end{equation}

Introducing dimensionless relative time parameter defined by,  \( \hat{t}= H_{o} \left( t-t_{0} \right)  \) , (87) can be written as,

\begin{equation}\label{eq88}
{\left(\frac{da}{d\hat{t}}\right)}^2=\ \frac{3}{u}{\mathit{\Omega}}_{m0}a^{2-u}+{\mathit{\Omega}}_{{\mathit{\Lambda}}_{C0}}a^2  
\end{equation}

Equation (88) gives the evolution of the scale factor for specific values of \textit{u} and  \(  \mathit{\Omega}_{m0} \) . In particular, we will plot our model Universe for the fitted parameter values obtained in section 6 and as well as for standard \textit{$ \Lambda$}CDM model with fitted parameter values from Union 2.1 analysis ~\cite{Suzuki}. The plot is shown in Figure~\ref{fig:scale factor plot}.   

\vspace{\baselineskip}
%%%%%%%%%%%%%%%%%%%%%%%%%%%%%%%%%%%%%%%%

\begin{figure*}

{\includegraphics[width=5.62in, keepaspectratio=true]{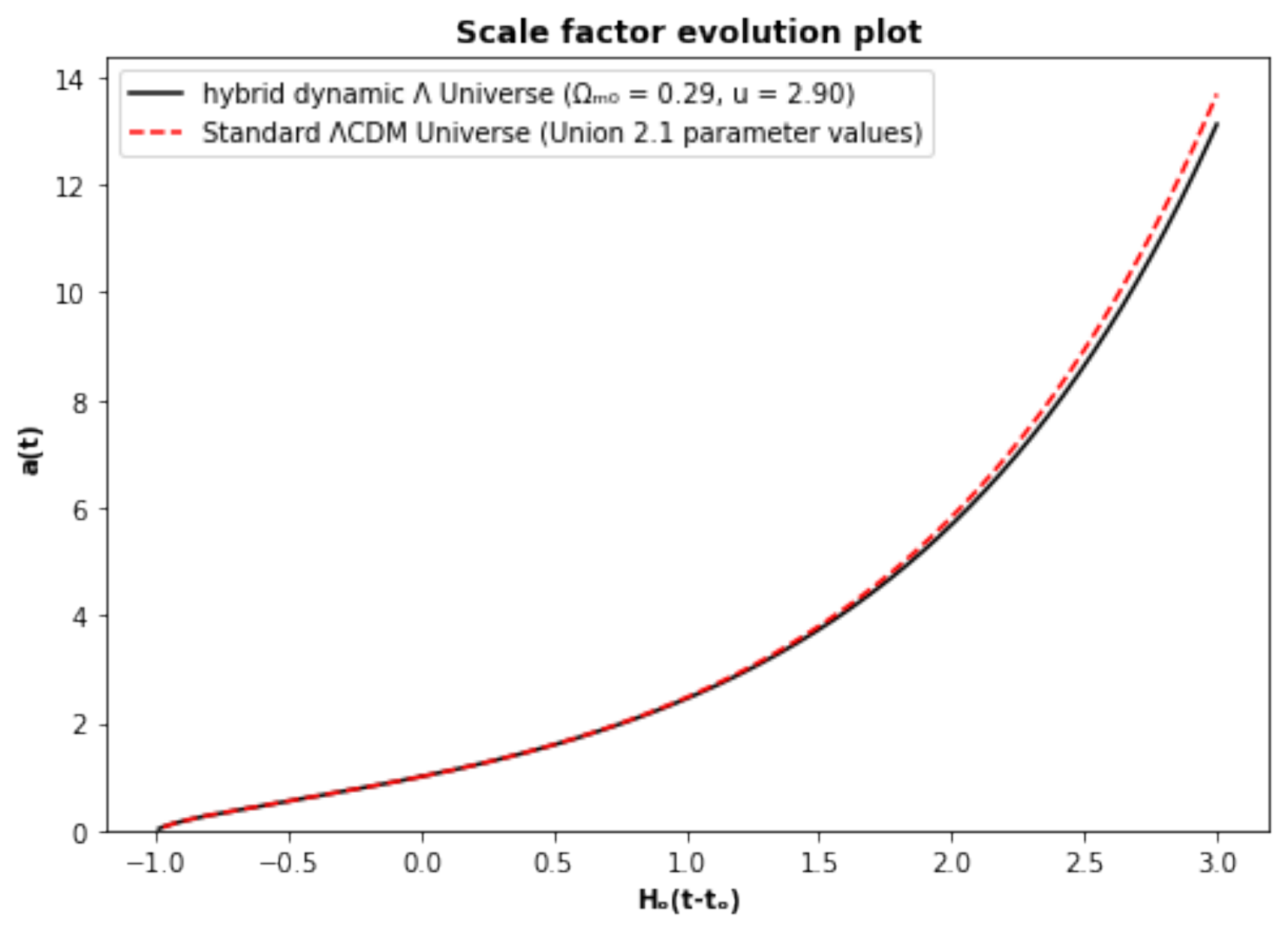}}

\caption{\textbf{\textit{Plot of scale factor with relative time parameter in units of Hubble time for standard $\boldsymbol{\mathit{\Lambda}}$CDM model and hybrid dynamic $\boldsymbol{\mathit{\Lambda}}$ model with Dust-Dark Energy (decaying) Universe.}}}
\label{fig:scale factor plot}
\end{figure*}

%%%%%%%%%%%%%%%%%%%%%%%%%%%%%%%%%%%%%%%%

%%%%%%%%%%%%%%%%%%%%%%%%%%%%%%%%%%%%%%%%

The side-by-side plot of scale factor evolution of standard model and hybrid dynamic \textit{$ \mathit{\Lambda}$} model indicates that the characteristics of the two types of models are similar and are almost indistinguishable at present epoch. However, there is a visible splitting between the two curves at future epochs showing that our model universe evolves slowly compared to standard model. However, such marginal deviation of scale factor evolution from \textit{$ \mathit{\Lambda}$}CDM case is unlikely to have any direct observational implication, specially at current epoch. 

\subsection{Deceleration parameter and transition redshift}
In section 4, it was pointed out that the deceleration parameter in hybrid dynamic \textit{$ \mathit{\Lambda}$} model exhibits a sign change reflecting transition of the Universe from decelerating to accelerating phase which is necessary for a realistic Universe. The present-day value of deceleration parameter and transition redshift can be easily derived for our model, from equation (75b) and (76), with the best fit parameter values obtained in section 6 as,

\[q_0=\ -0.56     ;  z_t=0.76\]

The plot of deceleration parameter as a function of redshift is shown in Figure~\ref{fig:deceleration parameter plot} both for our model and standard model. Once again, the two curve bear similarity in characteristics and the transition point for the two curves is so close that they cannot be distinguished from the graph (transition redshift quoted in Union 2.1 paper ~\cite{Suzuki} for standard model is  \( z_{t}=0.75 \) \  which is a close match to our model). The present-day deceleration parameter when computed for \textit{$\mathit{\Lambda}$}CDM model with Union 2.1 values yields  \( q_{0}= -0.59 \)  which is also very close to the values in our model. The marginal difference in the values indicates that the onset of accelerating phase happens earlier at smaller value of scale factor which is logical because in our model Universe scale factor evolves slowly compared to the standard model, but the differences lie within 1$ \sigma $  error levels which makes it very difficult to distinguish the models from observations.

%%%%%%%%%%%%%%%%%%%% Figure/Image No: 4 starts here %%%%%%%%%%%%%%%%%%%%
\begin{figure*}

{\includegraphics[width=5.71in, keepaspectratio=true]{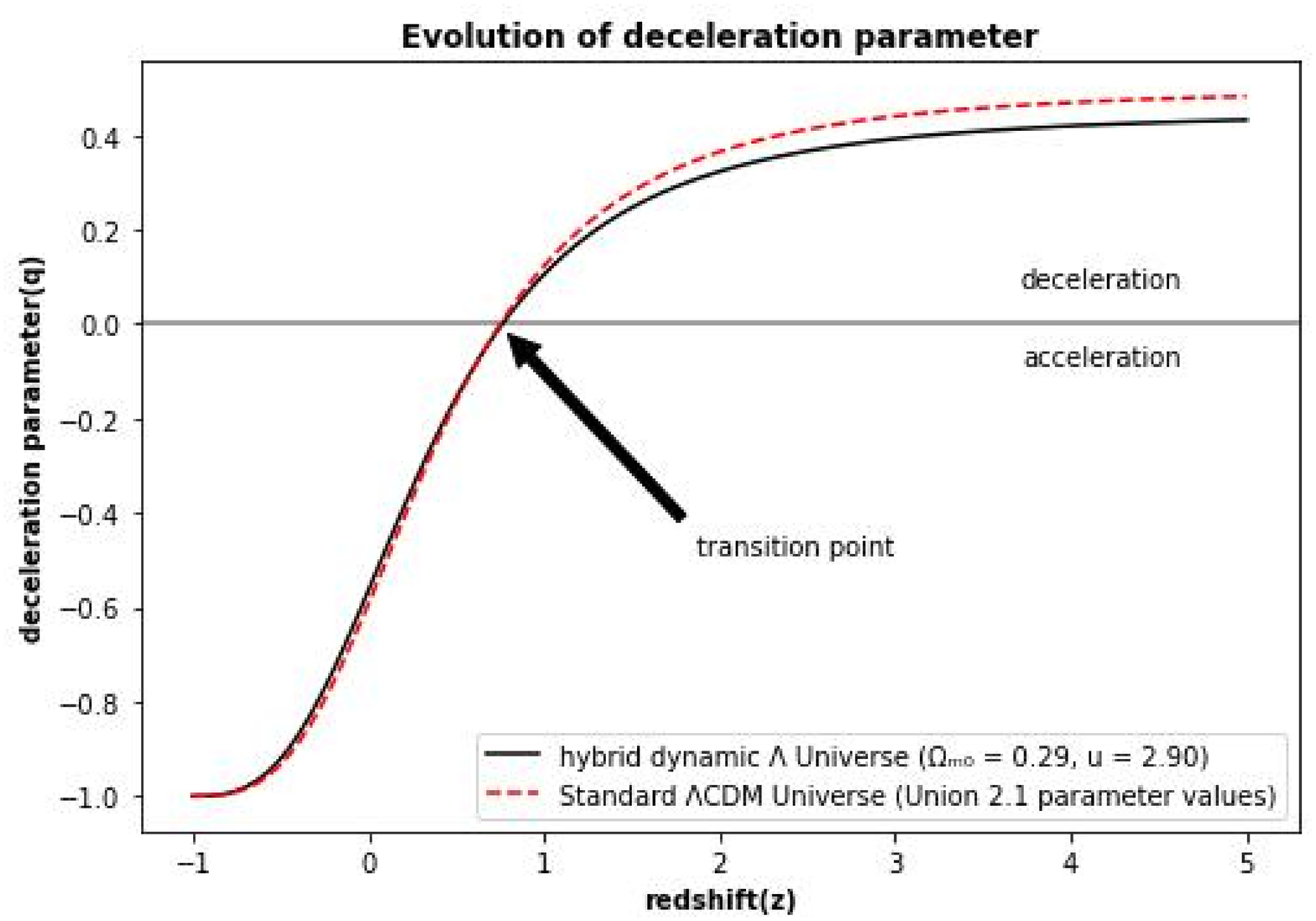}}

\caption{\textbf{\textit{Plot of deceleration parameter as a function of redshift for standard $\boldsymbol{\mathit{\Lambda}}$CDM model and hybrid dynamic $\boldsymbol{\mathit{\Lambda}}$ model with Dust-Dark Energy (decaying) Universe.}}}
\label{fig:deceleration parameter plot}
\end{figure*}

%%%%%%%%%%%%%%%%%%%%%%%%%%%%%%%%%%%%%%%%

\subsection{Age of the Universe and Cosmic Age problem}

It was mentioned in section 4 that a very important parameter in analysis of Cosmological models is the Cosmic Age parameter due to the associated $``$Cosmic Age Problem$"$  with this parameter. Most standard models of different eras have been bugged by this issue. It was previously thought that \textit{$ \mathit{\Lambda}$}CDM model is free from it since all the estimates of Cosmic Age in \textit{$ \mathit{\Lambda}$}CDM model from different surveys revolve around the values of 13.7 - 13.8 Gyr ~\cite{Hinshaw, Aghanim} which is above the lower limit on Age of the Universe (11 Gyr) set up by studying ages of Globular clusters ~\cite{Krauss}. However, recently some Globular Clusters were found which are older than the cosmic age of \textit{$\mathit{\Lambda}$}CDM model ~\cite{Ma, Wang2} with the oldest one (BO50) having an age of 16 GYR. It has brought the Cosmic Age problem back to limelight and till now there is no answer to this problem in the framework of standard \textit{$ \mathit{\Lambda}$}CDM Cosmology.

The phenomenological dynamic \textit{$\mathit{\Lambda}$} models stand in a very peculiar position in regard to Cosmic Age issue. The ages found in different model has been found to vary widely. While some models have Cosmic Age as low as 5.4 to 7.4 Gyr ~\cite{Overduin1}, others have Cosmic Age as high as 27.4 $ \pm $  5.6 Gyr ~\cite{Vishwakarma}. In our model, Cosmic Age can be estimated from equation (80) with the best-fit values obtained in section (6) which gives,

\[{Cosmic\ Age\ (t}_0)=13.93\ Gyr\] 

This is little higher than the estimates for standard \textit{$ \Lambda$}CDM models from various surveys (in comparison, Cosmic Age computed for \textit{$\mathit{\Lambda}$}CDM model using Union 2.1 parameter values yields 13.85 Gyr), but it is within 1$ \sigma $  error of the standard model values just like other parameters. However, the marginal increase in the value of Age is once again a reflection of the Universe evolving slowly compared to the standard model. Evidently, hybrid dynamic \textit{$\mathit{\Lambda}$} model with Dust and decaying Dark Energy does not solve the Cosmic Age problem but unlike many pure dynamic \textit{$\mathit{\Lambda}$} models, it does not suffer from high Age/Low Age issues. Rather it finds a value of Cosmic Age which is close to standard model and in this regard, it can be stated that the status of Cosmic Age problem in this model is at same footing with the standard model.

\subsection{Particle number density and creation rate parameter}

In equation (81), we obtained the expression for evolution of particle number density parameter with scale factor of the Universe. The present day value of particle number density for our model can be written in terms of dimensionless density parameter as,
\begin{equation}\label{eq89}
n_{m0} = \frac{\rho_{m0}}{m} = \frac{\mathit{\Omega}_{mo}\left(\frac{3H_{0}^2}{8\pi G}\right)}{m}
\end{equation}

For standard \textit{$\mathit{\Lambda}$}CDM model, a simillar expression can be obtained,

\begin{equation}\label{eq90}
n_{m0}^{\mathit{\Lambda}CDM} = \frac{\rho_{m0}^{\mathit{\Lambda}CDM}}{m} = \frac{\Omega_{mo}^{\mathit{\Lambda} CDM}\left(\frac{3H_{0}^2}{8\pi G}\right)}{m}
\end{equation}

where we have used the superscript \textit{$\mathit{\Lambda}$}CDM to represent the relevent quantities for \textit{$\mathit{\Lambda}$}CDM model. Since, present day matter density in \textit{$\mathit{\Lambda}$}CDM Universe is less than present day matter density in Universe with varying Cosmological Constant due to additional particles being created due to decay of vacuum energy, it follows  from (89) and (90) that,

\begin{equation}\label{eq91}
n_{m0} > n_{m0}^{\mathit{\Lambda}CDM}
\end{equation}
 
Equation (91) can be viewed as a general criteria that must hold when \textit{$\mathit{\Lambda}$} is a time-varying parameter as opposed to a constant one where decay of vacuum energy leads to a simultaneous creation of Dust particles. The present day particle creation rate parameter specific to the hybrid dynamic \textit{$\mathit{\Lambda}$} model can be estimated from equation (82a) or (82b) using the best fit values of Cosmological parameters which readily gives,

\[\Gamma_{mo}^{h = 0.7} = 0.227 \times 10^{-18} sec^{-1}\]

In figure~\ref{fig:creation rate parameter plot}, a plot of particle creation rate is shown as a function of redshift. 

\begin{figure*}
{\includegraphics[width=5.91in, keepaspectratio=true]{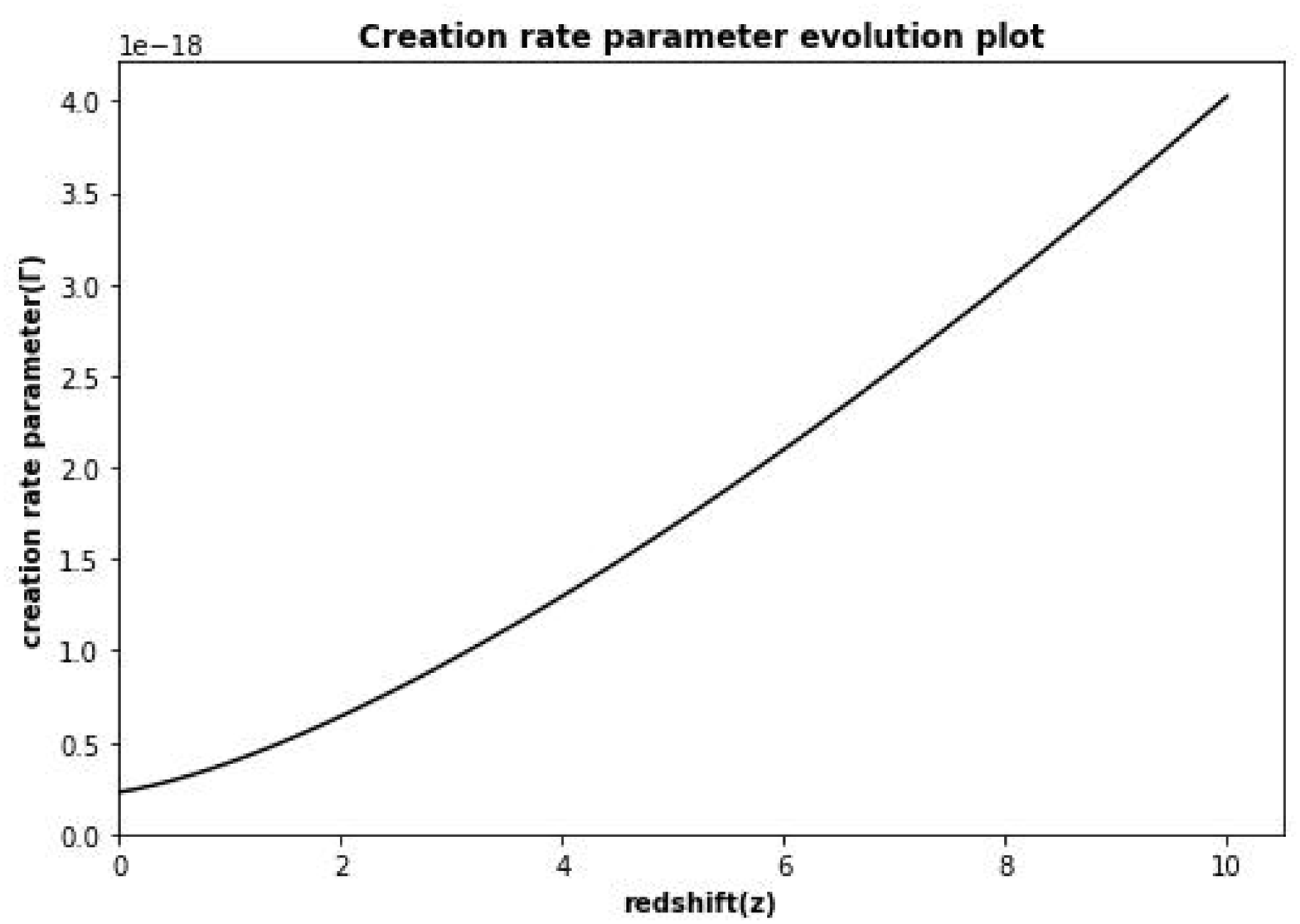}}
\caption{\textbf{\textit{Plot of creation rate parameter as a function of redshift for hybrid dynamic $\boldsymbol{\mathit{\Lambda}}$ model with Dust-Dark Energy (decaying) Universe.}}}
\label{fig:creation rate parameter plot}
\end{figure*}

\textit{$\mathit{\Lambda}$}(t) Cosmology thus leaves its signature through the creation of additional particles which alters the particle number density of the decaying vacuum product component compared to standard \textit{$\mathit{\Lambda}$}CDM model. In Cosmological models involving gravitationally induced continous particle creation from global curvature, particle production have an impact on structure formation in the Universe ~\cite{Nunes}. In an Universe with time-varying Cosmnological Constant, even though particle creation has a different mechanism, but it will also have a simillar impact on structure formation and will likely have observational consequences on weak lensing effect induced by structure formation. 

%%%%%%%%%%%%%%%%%%%% Table No: 1 starts here %%%%%%%%%%%%%%%%%%%%

\begin{table} [b]
\centering
\captionsetup{justification=centering}
\caption{Comparison table of Cosmological parameters}

\begin{tabular}{| c |c |c |} \hline 
Parameters & $\mathit{\Lambda}$CDM model (with Union 2.1 parameter values) & \makecell{Hybrid Dynamic \textit{$\mathit{\Lambda}$} model \\ (with Dust and decaying Dark Energy)} \\ \hline 
${\mathit{\Omega}}_{m0}$\textit{} & 0.271 & 0.29 \\ \hline 
$u$ & Fixed at 3.0 & 2.90 \\ \hline 
$q_0$ & -0.59 & -0.56 \\ \hline 
$z_t$ & 0.75 & 0.76 \\ \hline 
$t_0$ & 13.85 GYR & 13.93 GYR \\ \hline 
$\Gamma_{mo}$ & 0 & $0.227 \times 10^{-18} sec^{-1}$ \\ \hline
\end{tabular}
\label{table:parameter values}
\end{table}

%%%%%%%%%%%%%%%%%%%% Table No: 1 ends here %%%%%%%%%%%%%%%%%%%%

\section{Status of Cosmological problems}
As elaborated in section I, the motivation behind introduction of time varying Cosmological constant as opposed to a constant one revolves around addressing the fine tuning issues of standard \textit{$\mathit{\Lambda}$}CDM model viz. Cosmological Constant Problem and Coincidence Problem. In this section, we will check whether the issues are actually resolved or not. 

\subsection{Cosmological Constant problem}

Standard Cosmology is bugged by the Cosmological constant problem which refers to the issue of theoretically predicted value of quantum vacuum energy density  at Planck era being 120 orders of magnitude larger than the observed value of vacuum energy density in present era. To conveniently check the issue, we will introduce a new parameter, in the context of $\mathit{\Lambda}(t)$ Cosmology, namely, $``$vacuum energy density ratio parameter$"$ which will refer to the ratio of magnitude of vacuum energy densities at two distinct Cosmic times (or scale factors) in the history of evolution of the Universe. Mathematically, it can be expressed as,

\begin{subequations}
\begin{equation}\label{eq92a}
R_{\rho_{\mathit{\Lambda}}}(t_1, t_2) = \frac{\rho_{\mathit{\Lambda}}(t_{1})}{\rho_{\mathit{\Lambda}}(t_{2})} 
\end{equation}
\begin{equation}\label{eq92b}
R_{\rho_{\mathit{\Lambda}}}(a_1, a_2)= \frac{\rho_{\mathit{\Lambda}}(a_{1})}{\rho_{\mathit{\Lambda}}(a_{2})}
\end{equation}
\end{subequations}

In standard $\mathit{\Lambda}$CDM model, the ratio parameter is always unity since vacuum energy density is constant. In $\mathit{\Lambda}(t)$ Cosmology, the ratio parameter will take different values depending on choosen Cosmic times. Technically, the Cosmological Constant problem will be resolved in a specific $\mathit{\Lambda}(t)$ Cosmological model, if the model satisfies  the condition, $R_{\rho_{\Lambda}}(t_{Pl}, t_0) = R_{\rho_{\Lambda}}(a_{Pl}, a_0) ~ \sim 10^{120}$ where $t_{Pl}$ and $a_{Pl}$ denote the Cosmic time and scale factor corresponding to Planck era respectively while $t_0$ and $a_0$ represent the corresponding quantities in present era.

From equation (45) and the definitions of dimensionless density parameters, the expression of vacuum energy density in Planck era for hybrid dynamic $\mathit{\Lambda}(t)$ model with Dust and decaying Dark Energy can be written as,

\begin{equation}\label{eq93}
\rho_{\mathit{\Lambda}}(t_{Pl}) = \rho_{\mathit{\Lambda}}(a_{Pl}) = \left[\left(\frac{3-u}{u}\right)\mathit{\Omega}_{m0}a_{Pl}^{-u} + \mathit{\Omega}_{\mathit{\Lambda}_{C0}}\right] \frac{3H_{0}^2}{8\pi G}
\end{equation}

Simillarly, the expression of vacuum energy density in present era ($t_{0}$, $a_{0}$) is,

\begin{equation}\label{eq94}
\rho_{\mathit{\Lambda}}(t_{0}) = \rho_{\mathit{\Lambda}}(a_{0}) = \left[\frac{3-u}{u}\mathit{\Omega}_{m0}a_{0}^{-u} + \mathit{\Omega}_{\mathit{\Lambda}_{C0}}\right] \frac{3H_{0}^2}{8\pi G}
\end{equation}

Using equation (93) and (94), we can write down the vacuum energy density ratio parameter for vacuum energy density at Planck era and present era as, 

\begin{equation}\label{eq95}
R_{\rho_{\mathit{\Lambda}}}(a_{Pl}, a_0) = \frac{\rho_{\mathit{\Lambda}}(a_{Pl})}{\rho_{\mathit{\Lambda}}(a_{0})} = \frac{\frac{3-u}{u}\mathit{\Omega}_{m0}a_{Pl}^{-u} + \mathit{\Omega}_{\mathit{\Lambda}_{C0}}}{\frac{3-u}{u}\mathit{\Omega}_{m0}a_{0}^{-u} + \mathit{\Omega}_{\mathit{\Lambda}_{C0}}}
\end{equation}

For low values of Cosmic time, the first term in equation (87) will dominate and an approximate analytical solution of scale factor as a function of Cosmic time can be written as,

\begin{equation}\label{96}
a(t) = \left[H_0 \frac{u}{2}\sqrt{\left(\mathit{\Omega}_{m0}\frac{3}{u}\right)} t\right]^{\frac{2}{u}}
\end{equation}

From equation (96), the normalised scale factor corresponding to the end of Planck era ($t_{Pl} \sim 10^{-43} sec$) can be approximately obtained as  $a_{Pl} \sim  10^{-42} $.  Substituting the best fit values of Cosmological parameters and $a_{Pl}$ in (95), one obtains,
\[R_{\rho_{\mathit{\Lambda}}}(a_{Pl}, a_0) \approx 10^{120}\]

Henceforth, it can be concluded that hybrid dynamic \textit{$\mathit{\Lambda}$} model with Dust and Decaying Dark Energy is free from Cosmological Constant problem.\footnote{In a more realistic Cosmological model including radiation, the early Universe should be radiation dominated and the scale factor corresponding to Planck era should be approximately $a_{Pl} \sim 10^{-32}$. Assuming the expression of vacuum energy density is not affected by presence of radiation, it follows that $R_{\rho_{\mathit{\Lambda}}}(a_{Pl}, a_0) \approx 10^{91}$ This doesn't solve the traditional Cosmological Constant problem but still solves the Cosmological Constant problem when partial cancellation of boson and fermion vacuum energies is taken into account, which can decrease the quantum vacuum energy density at Planck era by upto $10^{36}$ orders of magnitude~\cite{Cheng} and bring down the discrepancy from $10^{120}$ to upto $10^{84}$ orders of magnitude.} 

\subsection{Coincidence problem}

Another vital issue faced by standard Cosmology is the coincidence problem which refers to the issue of vaccum energy density and matter density being of same order in the present era despite scaling differently invoking the question $``$why now?$"$. To evaluate the issue, we will use $``$time-dependent proximity parameter$"$ introduced by Egan and Lineweaver ~\cite{Egnan} which is defined as,

\begin{equation}\label{97}
r_p = min \left[\frac{\rho_{\mathit{\Lambda}}}{\rho_m}, \frac{\rho_m}{\rho_{\mathit{\Lambda}}}\right] 
\end{equation} 

Since the proximity parameter is a ratio of densities\footnote{proximity parameter is defined such that it is always the minimum of the two ratios}, it can also be expressed in terms of dimensionless density parameters as,

\begin{equation}\label{98}
r_p = min \left[\frac{\mathit{\Omega}_{\mathit{\Lambda}}}{\mathit{\Omega}_m}, \frac{\mathit{\Omega}_m}{\mathit{\Omega}_{\mathit{\Lambda}}}\right]
\end{equation} 

From the definition of proximity parameter (97), it is clear that if the two densities are exactly equal, then, $r_p = 1$ whereas if they differ by many orders of magnitude, then $r_p \sim 0$. In general if the two densities have same order of magnitude, then we will have $r_p\gg 0$. In present era, for $\mathit{\Lambda}$CDM model $ r_p  = 0.37 \gg 0$ (using Union 2.1 parameter values). However, in $\mathit{\Lambda}$CDM Cosmology, $\rho_{\Lambda}$ is a constant while $\rho_m$ is a time-varying parameter. Therefore, if we go back in Cosmic history in $\mathit{\Lambda}$CDM Universe, the proximity parameter will inevitably drift away from its present value towards $r_p \sim 0$ due to evolution of matter density. It implies that we coincidentally live in a special time in Cosmic history where the densities are of same order of magnitude and this is essentially the infamous coincidence problem. A Cosmological model will be free from Concidence problem if  $r_p \sim \mathcal{O}(1)$ is a general feature of the model holding for most of Cosmic history and is not a special occurance at present era. To check the situation in our model, we will write down the proximity parameter for our model in terms of present day values of density parameters which gives,

\begin{equation}\label{99}
 r_p (a) = min \left[\frac{\left(\frac{3-u}{u}\mathit{\Omega}_{m0}a^{-u} + \mathit{\Omega}_{\mathit{\Lambda}_{C0}}\right)}{\left(\mathit{\Omega}_{m0}a^{-u}\right)}, \frac{\left(\mathit{\Omega}_{m0}a^{-u}\right)}{\left(\frac{3-u}{u}\mathit{\Omega}_{m0}a^{-u} + \mathit{\Omega}_{\mathit{\Lambda}_{C0}}\right)}\right]
\end{equation}

Figure~\ref{fig:proximity parameter plot} shows the plot of proximity parameter as a function of scale factor for our model and standard $\mathit{\Lambda}$CDM model. 

\begin{figure*}
{\includegraphics[width=6.71in, keepaspectratio=true]{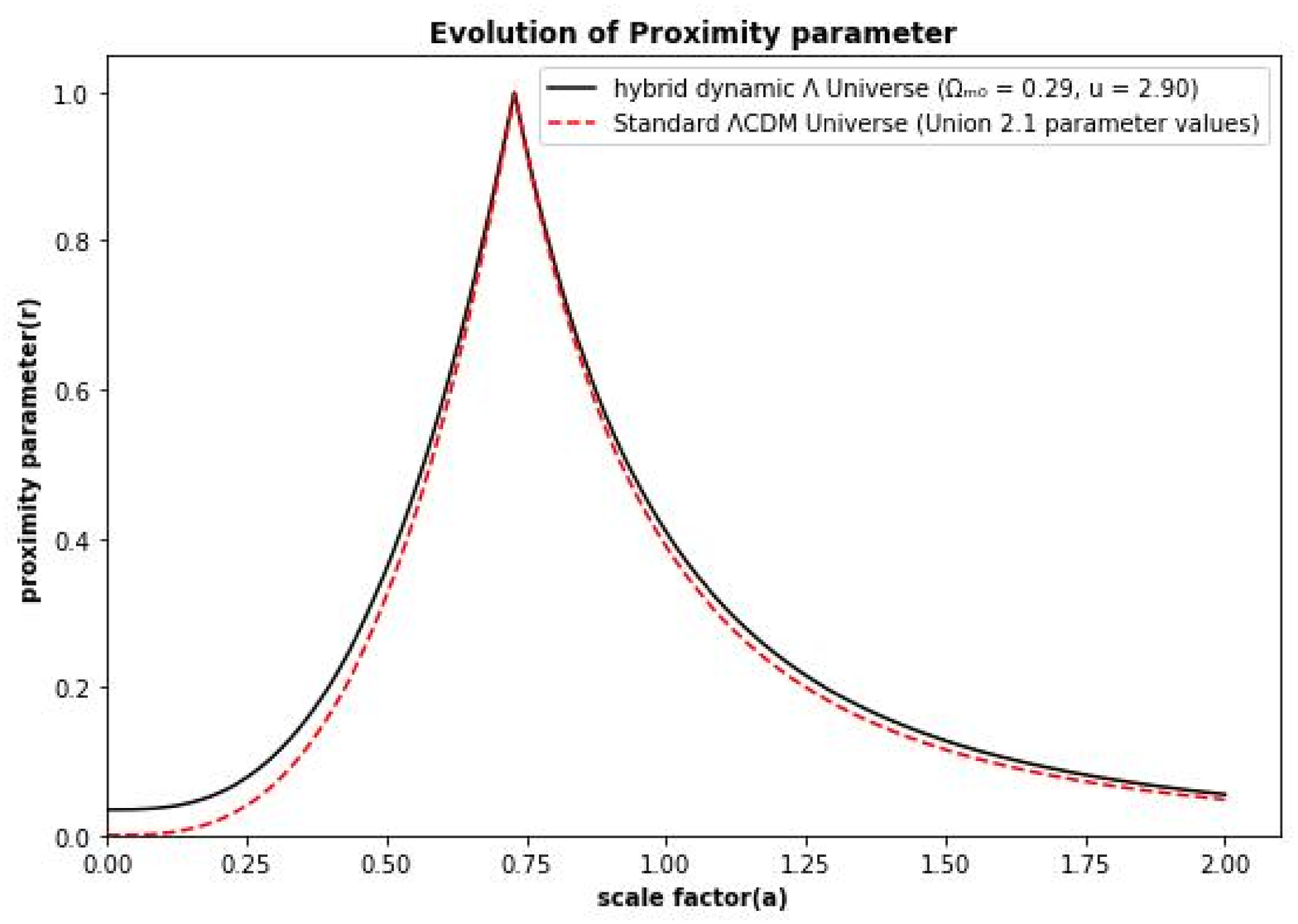}}
\caption{\textbf{\textit{Plot of proximity parameter as a function of scale factor for standard $\boldsymbol{\mathit{\Lambda}}$CDM model and hybrid dynamic $\boldsymbol{\mathit{\Lambda}}$ model with Dust-Dark Energy (decaying) Universe.}}}
\label{fig:proximity parameter plot}
\end{figure*}

The plots reveal that the pattern of evolution of proximity parameter in both the models is simillar which stays close to zero for most of cosmic history implying that the densities differ by many orders of magnitude but has a peak where the densities are comparable and the present era $a=1$ coincidentally lies near the peak. Therefore, it can be concluded that coincidence problem cannot be avoided by hybrid dynamic $\mathit{\Lambda}$ model. It is interesting to note that coincidence problem is often handled in literature by using some sort of tracker models where vacuum energy density evolves in a fashion simillar to evolution of matter and as a consequence the two densities maintain a constant ratio which removes the coincidence problem. A simillar feature is achieved in pure $\mathit{\Lambda}(t)$ Cosmological models as well where the constant component of Cosmological Constant is absent and evolution of vaccum energy density tracks the evolution of matter density. However, tracker models are unphysical in the sense that such Universes are either always accelerating or always decelerating. Addition of a constant term to $\mathit{\Lambda}(t)$ Cosmology makes the model physically viable by returning $\mathit{\Lambda}$CDM like behavior but along with it Coincidence problem returns as well! Since the coincidence problem in hybrid dynamic $\mathit{\Lambda}$ model originates due to the constant component of vacuum energy density, the coincidence problem can also be realised in a modified form,  $``$why the density of constant component of vacuum and matter density have same order in present epoch despite scaling differently?$"$. In figure~\ref{fig:density evolution plot}, the evolution of density parameters of $\mathit{\Lambda}$CDM model and hybrid dyanamic $\mathit{\Lambda}$ is shown to elaborate the simillarity in evolutions of density parameters for both the models which is eventually responsible for the simillar pattern of proximity parameter for both the models. 

\begin{figure*}
\subfigure[] {\includegraphics[width=5.71in, keepaspectratio=true]{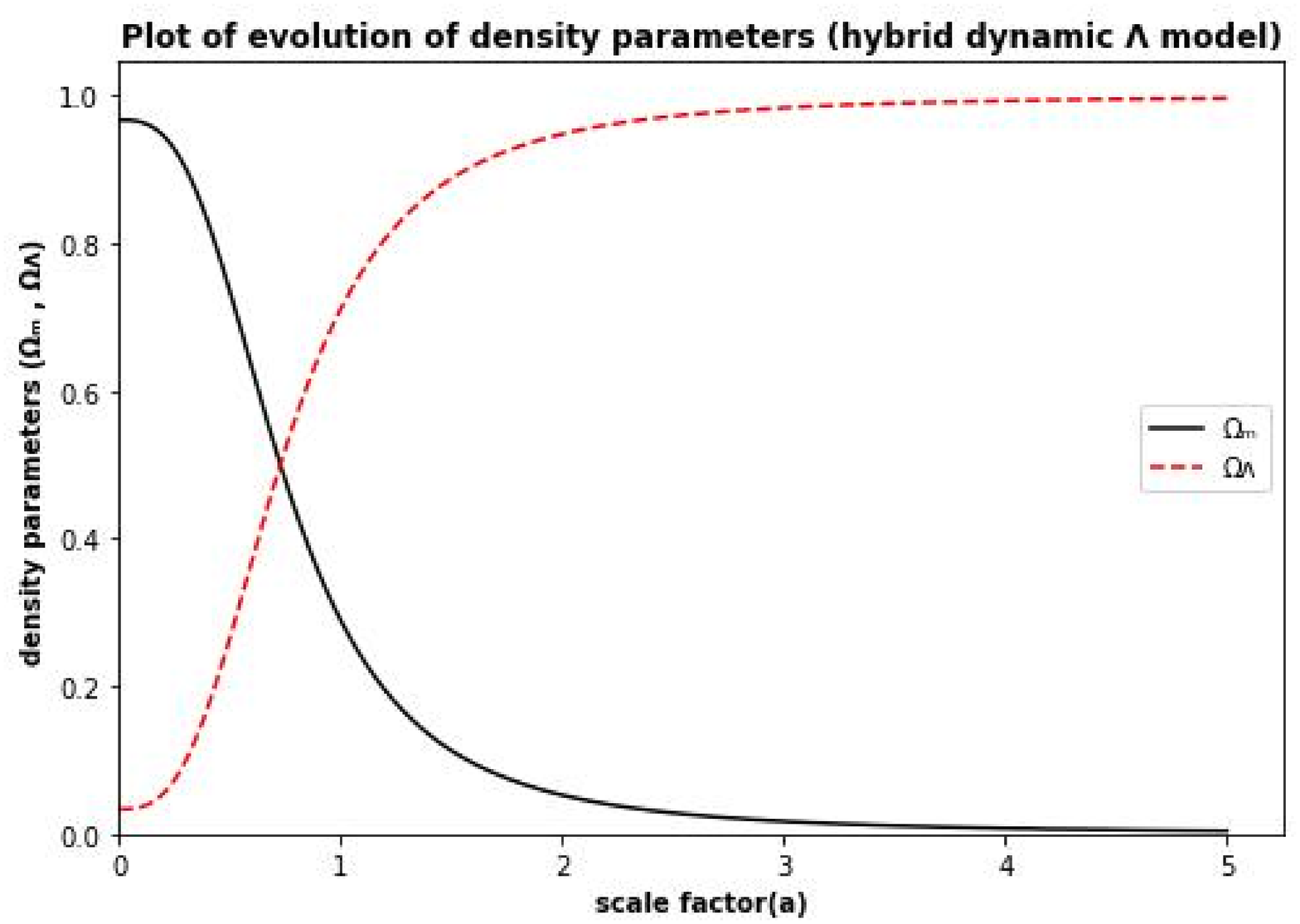}}
\subfigure[] {\includegraphics[width=5.71in, keepaspectratio=true]{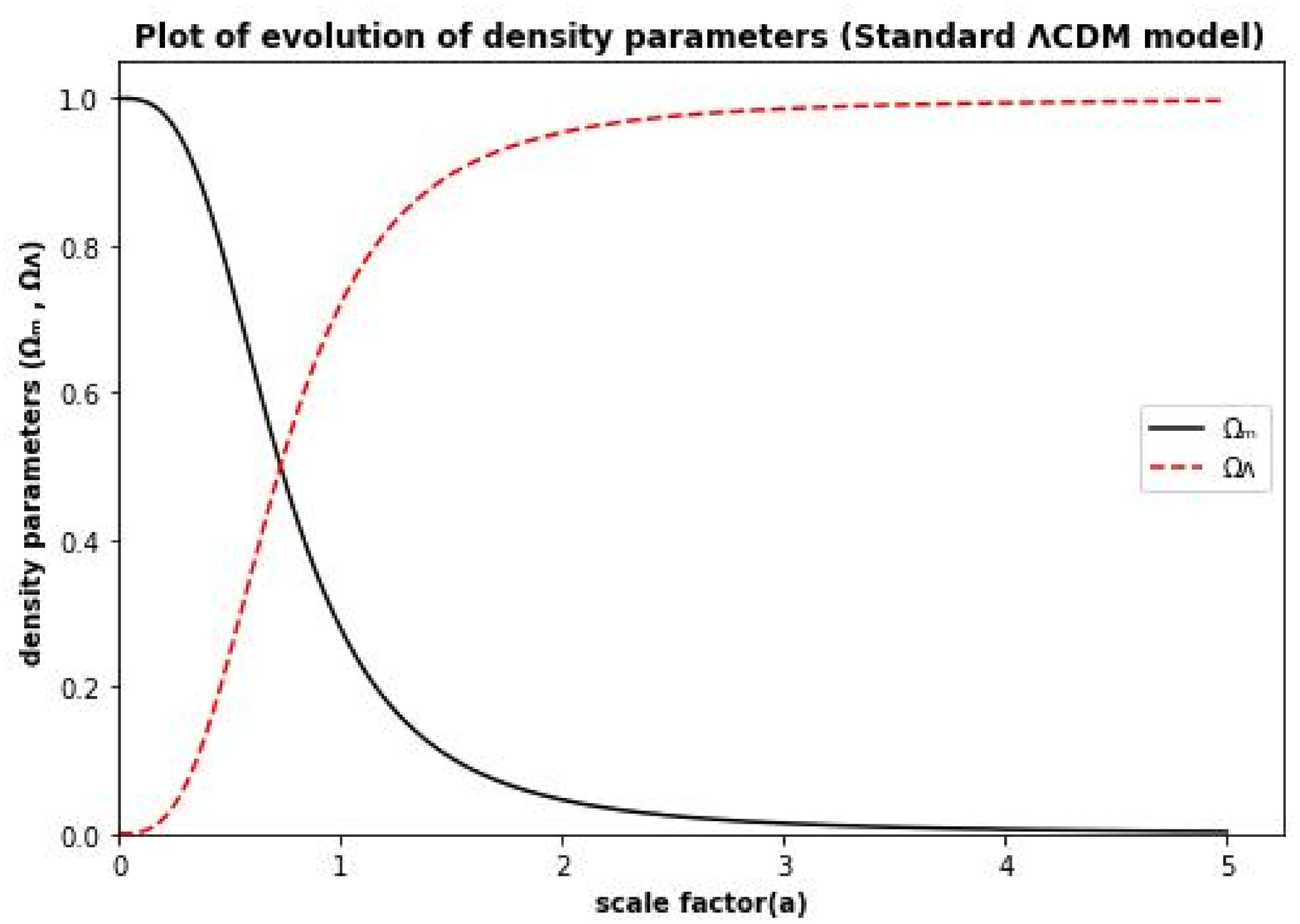}}
\caption{\textbf{\textit{Evolution of density parameters as function of scale factor for hybrid dynamic $\boldsymbol{\mathit{\Lambda}}$ model with Dust-Dark Energy (decaying) Universe (top plot) and standard $\boldsymbol{\mathit{\Lambda}}$CDM Universe (bottom plot).}}}
\label{fig:density evolution plot}
\end{figure*}    

To summarize, the hybrid dynamic nature of Cosmological Constant introduced in this work can successfully solve the Cosmological Constant problem but cannot remove the coincidence problem.  

\section{Discussions and Conclusion} \label{sec8}
In this work, we introduced a new type of phenomenological  \textit{$\mathit{\Lambda}$}(\textit{t})  model where  \textit{$\mathit{\Lambda}$}(\textit{t}) was taken to be a superposition of constant and variable components instead of being pure dynamic as usually assumed in most phenomenological models of variable  \textit{$\mathit{\Lambda}$}(\textit{t}) in literature. This was mainly done with the motivation of linking the phenomenological models with quantum models. However, we have not used any quantum theories or explanations in this work and our approach has been strictly phenomenological in all aspects. It can be thought of as a complementary approach which reciprocates metastable Dark Energy decay scenario from a classical phenomenological approach without resorting to quantum theories. 

For this investigation, we chose three expressions for variable component of  \textit{$\mathit{\Lambda}$}(\textit{t}) justified from dimensional analysis and solved for each of them separately. The solutions that we obtained for two-fluid model involving Dark Energy and another major component, showed that variable component of Dark Energy and the other major component decays at the same rate which rephrases the coincidence problem if the other major component is taken to be matter. Further, we found the three models to be equivalent which allowed us to devise a new parametrisation of Cosmological equations in terms of dilution rate of the components. The expression of deceleration parameter shows that the switch from the decelerating to accelerating phase can be readily obtained in our model which reflects that our model is physically sensible. In general, the Cosmological equations obtained in terms of dilution parameter showed resemblance to standard \textit{$\mathit{\Lambda}$}CDM model. In fact, the corresponding equations for standard model can be readily obtained just by substituting  \( u = 3 \)  in all the equations for Dust-Dark Energy (decaying) type hybrid dynamic \textit{$\mathit{\Lambda}$} model and by substituting  \( u = 3 \left( 1+ \omega  \right)  \)  along with  \(  \mathit{\Omega}_{0} =  \mathit{\Omega}_{m0} \)  for the general case. The standard \textit{$ \mathit{\Lambda}$}CDM model can thus be obtained as a special case for our model, which provides a ground for comparison and can be viewed as an added advantage of our model. 

Finally, on confronting the model with observations using 580 supernovae from Union 2.1 dataset, we found that for Dust-Dark Energy (decaying) Universe, the parameter estimates are  \(  \mathit{\Omega}_{m0}~= 0.29   \pm  0.03 \) \ ,   \( u~= 2.90   \pm  0.54 \) . The corresponding values of derived parameters were found to be,  \( q_{0} = - 0.56 \) \ ,   \( z_{t}~= 0.76 \) ,  \( t_{0} = 13.93 Gyr \) , \(\Gamma_{m0} = 0.227 \times 10^{-18} sec^{-1}\). The plots of variation of scale factor and deceleration parameter with redshift showed similar pattern with standard model, but the Universe seemed to evolve little bit slowly compared to the standard case which is also reflected by the marginal deviation of the parameter values in comparison to standard case. Table~\ref{table:parameter values} summarizes the Cosmological parameters of standard $\mathit{\Lambda}$CDM model and hybrid Dynamic \textit{$\mathit{\Lambda}$} model and provides a quick comparison. 

The closeness of the model parameters estimates with standard \textit{$ \mathit{\Lambda}$}CDM model means that it is difficult to distinguish the model from \textit{$ \mathit{\Lambda}$}CDM model from observational studies. However, on the contrary, it can also be stated that any observational results which favours \textit{$ \mathit{\Lambda}$}CDM model, will also favour our model since our model retains most of the conclusions drawn from \textit{$ \mathit{\Lambda}$}CDM model. Additionally, our model solves one of the alarming issues of modern Cosmology - $``$cosmological constant problem$"$. The coincidence problem however cannot be evaded in our model. The Cosmic Age increases marginally which does not solve the Cosmic Age problem. However, the uncertainties on the parameters are high and instead of taking the best-fit values, if we roughly take a point at the boundary of the 1$ \sigma $  contour in figure 2, say,  \(  \left(  \mathit{\Omega}_{m0}=0.31, u=2.475 \right)  \) [shown by a mark in Figure~\ref{fig:Corner Plots}b], then we immediately get a cosmic age of 15.31 GYR which can accommodate all the old globular clusters in ~\cite{Ma,Wang2} except the oldest one and is very close to solving the Cosmic Age Problem! We are not claiming this to be true since it is very absurd for the dilution rate to deviate that much from the standard case but we are merely pointing out the necessity of reducing the uncertainty. Henceforth, from observational or computational point of view, an immediate extension of the work can be to perform joint statistical analysis using different methods and more datasets so that the tighter constraints can be imposed on the model parameters. From a theoretical point of view, it will also be interesting to see how the hybrid dynamic \textit{$ \mathit{\Lambda}$} model behaves when combined with a varying Gravitational Constant (\textit{G}) scenario, an idea that comes out from Dirac's large number hypothesis (LNH) ~\cite{Dirac}.  

We conclude this work by saying although our model is phenomenological in nature, but it gives very sensible conclusions about the Universe and have the potential to become a precursor to a more concrete theory that might be developed in the future.

\section*{Acknowledgments}

I would like to express my heartful gratitude towards my supervisor Dr. Bryan Carpenter, School of Computing, University of Portsmouth for his help and guidance in completing this work. Working under him has been a wonderful experience. I am also thankful to my project advisors Professor Robert Crittenden,  Institute of Cosmology and Gravitation, University of Portsmouth, and Dr Thomas Kecker, School of Mathematics $\&$  Physics, University of Portsmouth for their valuable advice and suggestions. I would also like to thank our course co-ordinator Dr. Alice Good for her general guidance in completing the course. Lastly, I would like to extend my gratitude to the anonymous referees for their recommendations and suggestions which greatly helped to improve the quality of the manuscript.

\section*{Funding}

This research did not receive any specific grant from funding agencies in the public, commercial, or not-for-profit sectors. However, since the project was a part of the MRES TECHNOLOGY program of University of Portsmouth, it benefitted from general resources provided by the University for research students.

\appendix

\section{Time varying Cosmological Constant and variable mass particles (VAMP)}

In section IV.F, we developed the framewok of particle creation effects in our model with the presupposition that mass of particles of other major component is invariant. In this section, being inspired by Variable Mass Particle (VAMP) Cosmology~\cite{Anderson, Leon}\footnote{see the references within ~\cite{Leon} for an account of different works done VAMP models}, we will expand our work to include the case where mass of the other major component is variable. In the context of our model, VAMP scenario stands for the situation where the decay of vacuum energy do not create new particles, rather the decay of vacuum is associated with variation (increase) of mass of the particles of the decaying vacuum product component. In this case, the equations developed in section IV.F and VII.D will not hold and the conservation equation (6) written in terms of mass and number density will take the form,

\begin{equation}\label{eqA1}
n\dot{m}\ + m\dot{n}\ \ + 3\frac{\dot{a}}{a}mn \left( 1 + \omega \right) = - \dot{\rho_{\mathit{\Lambda}}}
\end{equation}

equation (A1) will replace equation (64) in this case. Since additional particles are not created due to decay of vacuum energy, number density of the other major component will follow the standard energy conservation equation given by,

\begin{equation}\label{eqA2}
\dot{n} + 3n\frac{\dot{a}}{a}\left(1+\omega\right) = 0
\end{equation}

Substituing equation (A2) in equation (A1) and simplifying, we have,

\begin{equation}\label{eqA3}
\dot{m}\ - 3m\frac{\dot{a}}{a}(1 + \omega) +\ 3\frac{\dot{a}}{a}m \left( 1 + \omega \right) =  - m\frac{\dot{\rho_{\mathit{\Lambda}}}}{\rho}
\end{equation}
 
Using expression of $\rho$ and $\rho_\mathit{\Lambda}$ from equation (45), equation (A3) can be written as,

\begin{equation}\label{eqA4}
\dot{m} -3m\frac{\dot{a}}{a}\left(1 + \omega\right) + mu\frac{\dot{a}}{a} = 0
\end{equation}

Equation (A4) can be readily solved to obtain,

\begin{equation}\label{eqA5}
m = m_0a^{3(1 + \omega) -u}
\end{equation}

where $m_0$ denotes the present day mass of the particles of the other major component. Equation (A5) gives the variation of mass of the particles of decaying vacuum product component due to decay of vacuum energy. When the other major component is dust with $\omega = 0$, equation (A5) will take the form,

\begin{equation}\label{eqA6}
m_m = m_{m0}a^{3 - u}
\end{equation}

The variation of mass of product component is an alternative scenario to the particle production picture. This alternative scenario doesn't change the variation of density parameters in the model. Hence, the entire Cosmology with time-varying \textit{$\mathit{\Lambda}$} developed throughout this work, except particle creation effects, will remain unaltered under such considerations.   

\section{Testing the model with Pantheon supernova compilation}

Here we present an updated result by fitting the model against latest supernovae data from Pantheon compilation~\cite{Scolnic}. Pantheon catalogue~\cite{dataset2} comprises of a dataset 1048 supernovae as $(m_{obs},z)$ pairs. The apparent magnitude reported in Pantheon dataset is actually the corrected apparent magnitude obtained after determining and adjusting all the nuissace parameters (except absolute Magnitude) by BEAMS with Bias Corrections (BBC) method~\cite{Kessler}. Since the publicly available Pantheon dataset is presented in ~\cite{dataset2} without setting up any pre-determined value for absolute magnitude, one can use any one of the independently measured value of Hubble Constant from observations along with Pantheon sample. The corresponding value of absolute magnitude and its uncertainty can be determined from the following equations used in~\cite{Raveri, Benevento},

\begin{equation}\label{eqB1}
M = 5log_{10}\frac{H_0^{meas}}{H_0^{fid}} + M^{fid}
\end{equation}

\begin{equation}\label{eqB2}
\sigma_M = \frac{5}{ln10}\frac{\sigma_{H_0^{meas}}}{H_0^{meas}}
\end{equation}

where $H_0^{fid} $ and $M^{fid}$ are the fiducial values of Hubble Constant and absolute magnitude relevent to Pantheon sample\cite{Raveri, Benevento}.\footnote{the fiducial values corresponds to the values used for training Pantheon supernova sample. Fiducial value of Hubble Constant in training of Pantheon sample was taken to be 70 whereas fiducial value of absolute magnitude can be extracted from trial Cosmological fits with $H_0^{fid} = 70$. We have readily adopted the fiducial values from ~\cite{Raveri, Benevento}. Interested readers may see ~\cite{Raveri} for details of the extraction process.} $H_0^{meas}$ and $\sigma_{H_0}^{meas}$ are the measured value of Hubble Constant and uncertainty respectively which will be utilised for Cosmological fitting.

Currently there is an ongoing tension in the value of Hubble Constant between its value determined from CMB measurements  $(67.4 \pm 0.05)$~\cite{Aghanim} and its value determined from local Supernova measurements ($74.03 \pm 1.42$)~\cite{Riess2}. We will take a conservative approach and use the model-independent measurement of Hubble constant  $H_0 = 69.8 \pm 0.8$ from Carnegie-Chicago Hubble Program~\cite{Freedman} which sits in the middle of the Hubble tension. The consequent value of absolute magnitude and its uncertainty determined from equations (B1) and (B2) are $M = -19.34621 \pm 0.02489$. Plugging in all these together, the best fit values of model parameters obtained are,

\[\mathit{\Omega}_{m0} = 0.31 \pm 0.02 ~~~~~~     ; ~~~~~~~    u = 2.92 \pm 0.34 \]
(errors reported are 1$ \sigma$).

The Hubble diagram and Corner plots are presented in figure \ref{fig:Hubble Pantheon plot} and figure \ref{fig:Corner Pantheon plot} respectively. 
\begin{figure*}
\centering
{\includegraphics[width=7.71in, keepaspectratio=true]{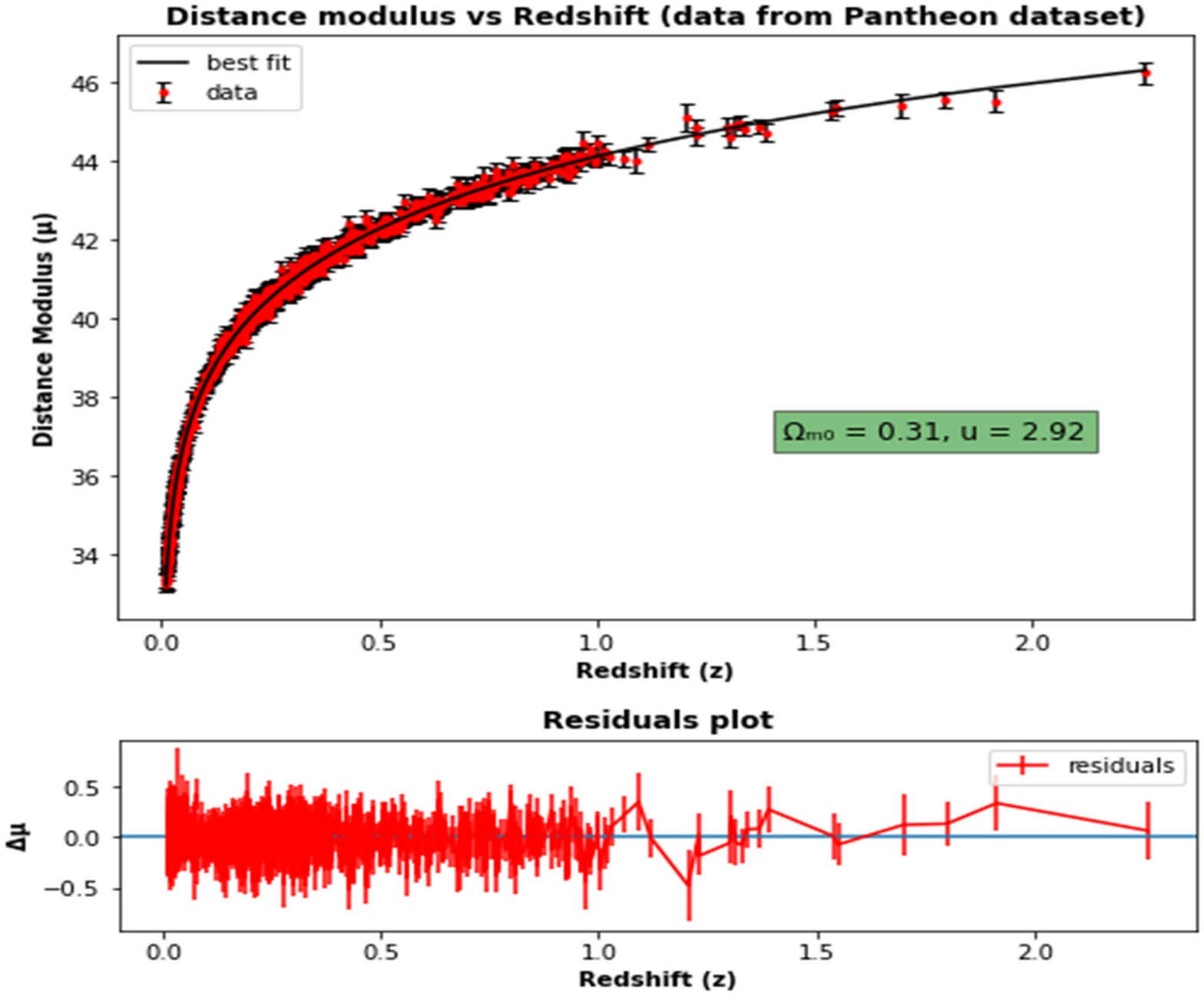}}
\caption{\textbf{\textit{Hubble diagram for hybrid dynamic $\boldsymbol{\mathit{\Lambda}}$ model with Dust-Dark Energy (decaying) Universe (top plot) and residuals (bottom plot) using Pantheon supernova sample.}}}
\label{fig:Hubble Pantheon plot}
\end{figure*}

\begin{figure*}
\centering
{\includegraphics[width=5.71in, keepaspectratio=true]{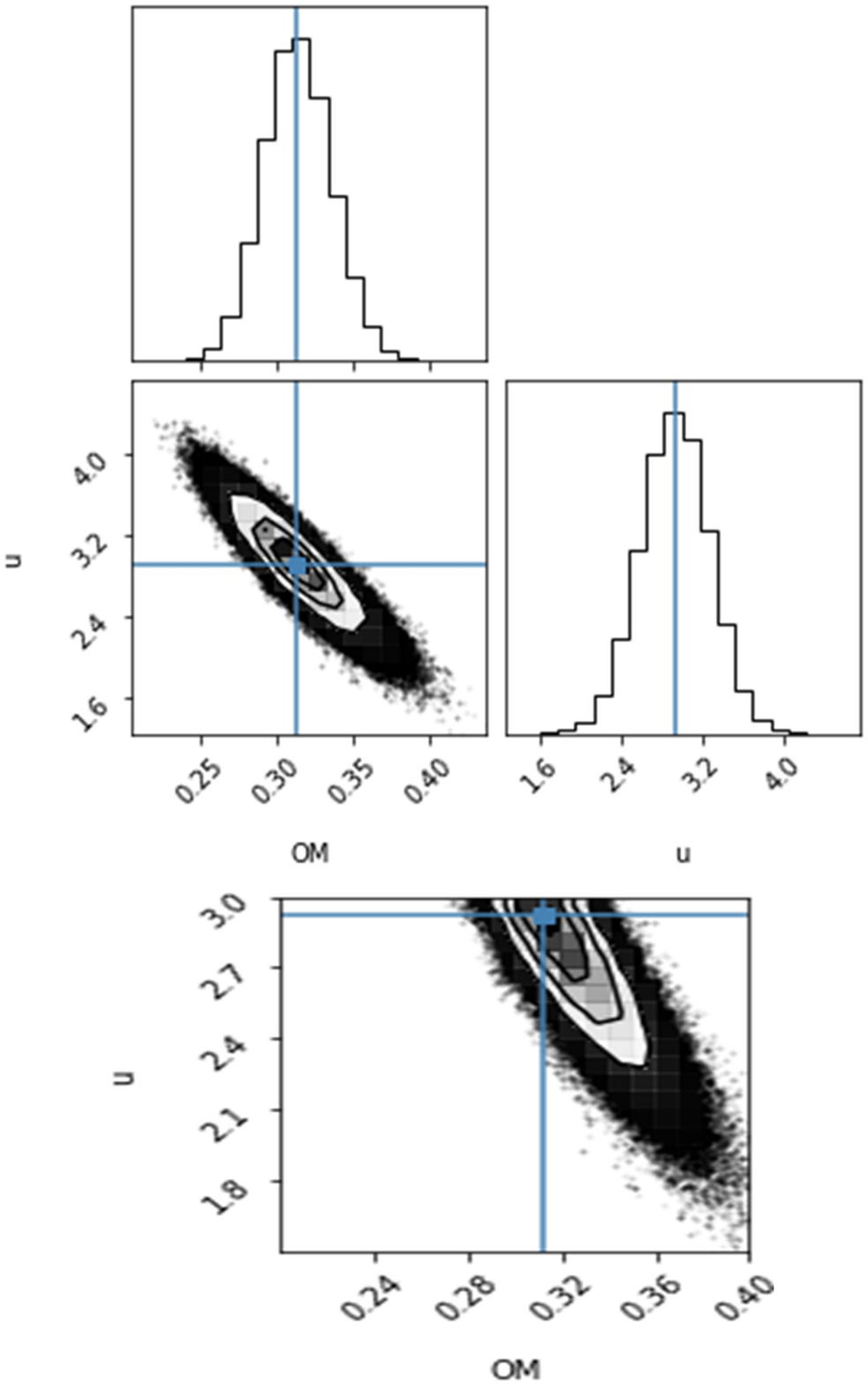}}
\caption{\textbf{\textit{The top plots are the Corner plot output showing results of our markov chain monte carlo parameter estimation for hybrid dynamic $\boldsymbol{\mathit{\Lambda}}$ Cosmological model with Dust-Dark Energy (decaying) Universe, using Pantheon supernova sample. Bottom plot is the zoomed in image of parameter contours which has been terminated at u = 3 to represent realistic scenario. Note- The label OM in the above figures is equivalent to $\boldsymbol{\mathit{\Omega}_{m0}}$}}}
\label{fig:Corner Pantheon plot}
\end{figure*}

Fitting with the larger Pantheon supernova sample have decreased the uncertainty in parameters. The best-fit values have shifted as well but the shifting is marginal and it is not expected to change any of the conclusions of this work significantly. 

\end{document}